\documentclass[english,aps,superscriptaddress,eqsecnum]{revtex4}
\usepackage[T1]{fontenc}
\usepackage[latin9]{inputenc}
\usepackage{babel}
\usepackage{verbatim}
\usepackage{amsbsy}
\usepackage{amstext}
\usepackage{amssymb}
\usepackage{esint}
\usepackage{color}
\usepackage{ulem}
\usepackage[bookmarks=true,colorlinks,linkcolor=blue,urlcolor=blue,citecolor=red]{hyperref}
\hypersetup{
 pdfauthor={Xin-li Sheng}}

\makeatletter
\@ifundefined{textcolor}{}
{%
 \definecolor{BLACK}{gray}{0}
 \definecolor{WHITE}{gray}{1}
 \definecolor{RED}{rgb}{1,0,0}
 \definecolor{GREEN}{rgb}{0,1,0}
 \definecolor{BLUE}{rgb}{0,0,1}
 \definecolor{CYAN}{cmyk}{1,0,0,0}
 \definecolor{MAGENTA}{cmyk}{0,1,0,0}
 \definecolor{YELLOW}{cmyk}{0,0,1,0}
}

\makeatother

\begin{document}

\title{Kinetic theory with spin: From massive to massless fermions}

\author{Xin-Li Sheng}
\affiliation{Interdisciplinary Center for Theoretical Study and Department of
Modern Physics, University of Science and Technology of China, Hefei,
Anhui 230026, China}

\author{Qun Wang}
\affiliation{Interdisciplinary Center for Theoretical Study and Department of
Modern Physics, University of Science and Technology of China, Hefei,
Anhui 230026, China}


\author{Xu-Guang Huang}
\affiliation{Physics Department and Center for Particle Physics and Field Theory, Fudan University, Shanghai 200433, China}
\affiliation{Key Laboratory of Nuclear Physics and Ion-beam Application (MOE), Fudan University, Shanghai 200433, China}

\begin{abstract}

We find that the recently developed kinetic theories with spin for massive and massless fermions are smoothly connected.
By introducing a reference-frame vector, we decompose the dipole-moment tensor into electric and magnetic dipole moments. We show that the axial-vector component of the Wigner function contains a contribution from the transverse magnetic dipole moment which accounts for the transverse spin degree of freedom (DOF) and vanishes smoothly in the massless limit. As a result, the kinetic equations, describing four DOF for massive fermions, becomes smoothly the chiral kinetic equations describing two DOF in the massless limit. We also confirm the small-mass behavior of the Wigner function by explicit calculation using a Gaussian wave packet.
\end{abstract}

\maketitle

\section{Introduction}

In non-central heavy-ion collisions, a large orbital angular momentum (OAM) \cite{Liang:2004ph,Betz:2007kg,Becattini:2007sr} as well as
a very strong electromagnetic field
\cite{Skokov:2009qp,Voronyuk:2011jd,Deng:2012pc,McLerran:2013hla,Tuchin:2013apa,Deng:2014uja,Li:2016tel}
are generated. Part of the OAM is transferred to the hot and dense matter
or the quark-gluon plasma in the form of vorticity fields \cite{Deng:2016gyh,Jiang:2016woz,Pang:2016igs,Deng:2020ygd}
and leads to a global spin polarization perpendicular to the reaction plane \cite{Liang:2004ph,Voloshin:2004ha,Fang:2016vpj,Gao:2007bc,Huang:2011ru,Becattini:2013fla,Zhang:2019xya}.
The global spin polarization of $\Lambda$ hyperons has been observed
by STAR collaboration in Au+Au collisions at $\sqrt{s_{NN}}=7.7-200$ GeV
\cite{STAR:2017ckg,Adam:2018ivw}, see, e.g. Refs. \cite{Wang:2017jpl,Huang:2020xyr,Becattini:2020ngo} for recent reviews.
The interaction between the strong magnetic field and fermion spin leads to
the chiral magnetic effect (CME) \cite{Vilenkin:1980fu,Kharzeev:2007jp,Fukushima:2008xe},
which can probe the topological fluctuation of quantum chromodynamics vacuum. The search for the CME is one of the major efforts
in experiments of heavy ion collisions; see, e.g. Refs. \cite{Kharzeev:2015znc,Huang:2015oca,Hattori:2016emy,Zhao:2019hta,Li:2020dwr,Liu:2020ymh} for reviews.

For massless fermions with definite helicity,
the chiral kinetic theory (CKT) is a useful tool to
describe the chiral effects in phase space \cite{Son:2012wh,Son:2012zy,Stephanov:2012ki,Gao:2012ix,Chen:2012ca,Hidaka:2016yjf,Huang:2018wdl,Gao:2018wmr,Liu:2018xip,Hidaka:2017auj,Gao:2017gfq,Carignano:2018gqt,Lin:2019ytz,Huang:2018aly}.
The Lorentz invariance for chiral fermions is proved to be non-trivial:
the side-jump effect appears to ensure the conservation of the total angular momentum
in binary collisions \cite{Chen:2014cla,Chen:2015gta,Gao:2018jsi}. Recent numerical
simulation \cite{Liu:2019krs} shows that the side-jump effect
may provide a possible explanation of the puzzle of the $\Lambda$'s local spin
polarization \cite{Becattini:2017gcx,Adam:2019srw}.

In reality, all quarks have masses. Although the masses of up and down quarks are small compared to the typical temperature of the quark-gluon plasma, the mass of strange quark is not.
The massive strange quarks play an essential role in the $\Lambda$'s spin polarization
as well as in the spin alignments of $\phi$ or $K^{\ast0}$ mesons described by the $00$-component of the spin density matrix \cite{Liang:2004xn,Yang:2017sdk,Sheng:2019kmk}.
Therefore a kinetic theory for massive fermions
with spin, also called spin kinetic theory, is required to describe the spin evolution of massive quarks in phase space.
Such a theory was constructed many years ago for non-relativistic dilute spinful gases \cite{Hess:1966,Hess:1968} and has recently been formulated for relativistic Dirac fermions using the covariant
Wigner functions \cite{Gao:2019znl,Weickgenannt:2019dks,Hattori:2019ahi,Liu:2020flb,Yang:2020hri}
and equal-time Wigner functions \cite{Wang:2019moi}.
It can also be constructed in the worldline formalism \cite{Mueller:2017arw,Mueller:2017lzw}.

However, there are fundamental differences between symmetries of massive and massless
fermions. In 1939, Wigner proposed the concept of the little group \cite{Wigner:1939cj},
the group that leaves a particle's four-momentum invariant.
For a massive particle, the little group is the rotational group $O(3)$,
which is associated with the spin in the particle's rest frame.
Wigner also showed that the little group
for the massless particle is the two-dimensional Euclidean group $E(2)$.
The rotational degree of freedom of $E(2)$ corresponds to the helicity,
while the two translational degrees of freedom correspond to the gauge
symmetry of the massless particle \cite{Wigner:1939cj,Kim:1986gq,Kim:1989wt}.
It can be proved that the $E(2)$ group can be obtained as the infinite-boost
limit or massless limit of the $O(3)$ group \cite{Kim:1986gq}. However, how the kinetic theories for massive fermions and massless fermions are connected under such a limiting process is unclear. There are proposals to make a smooth transition between the two kinetic theories
 \cite{Weickgenannt:2019dks,Hattori:2019ahi},
but these proposals are based on the assumption that the spin polarization or dipole-moment tensor
for massive fermions can be smoothly reduced to its massless forms when $m\rightarrow0$. This assumption has not been justified.

In this paper, we propose a kinetic theory
with spin in the Wigner function formalism
that can smoothly transit between the massive and massless cases. The main idea is to project out the transverse-spin contributions in the kinetic theory and to show that they smoothly vanish in the massless limit. In such a way, two of the four kinetic equations in the massive case become irrelevant in the massless limit, leaving the other two to constitute the CKT that describes the vector- and axial-charge distributions.
The paper is organized as follows.
In Sec. \ref{sec:Wigner-function-at}, we briefly review the Wigner
function formalism for massive fermions up to the linear order in $\hbar$.
Then in Sec. \ref{sec:Reference-frame-dependence}, to connect with
the massless case, we introduce a reference frame in the massive case.
The dipole-moment tensor and spin-polarization vector are expressed in terms of
reference-frame dependent quantities such as the fermion number density,
the magnetic dipole moment, and the axial-charge density.
In Sec. \ref{sec:Wigner-function-for}, the ensemble of particle states of
the Wigner function is replaced by the one-particle wave-packet state
with finite momentum and space-time dispersion. In this case
we derive explicit expressions for the fermion number density,
the magnetic dipole moment, and the axial-charge density and obtain their small-mass behaviors. In Sec. \ref{sec:Mass-correction-to}, the Wigner-function components
as well as the kinetic equations are written in forms with mass corrections explicitly singled out.
Turning off the mass corrections, we can recover the massless case smoothly.
A summary of the results is given in the final section.

Throughout this paper, we use the units $c=k_{B}=1$ but keep the reduced
Planck's constant $\hbar$ explicitly. The electromagnetic potential
is labeled by $\mathbb{A}^{\mu}$ with the electric charge being absorbed.
This means that we can recover the electric charge $Q$ of the fermion
by replacing $\mathbb{A}^{\mu}\rightarrow Q\mathbb{A}^{\mu}$. We consider the background spacetime to be Minkowskian, but all the calculations can be similarly carried out in curved spacetime as well (The kinetic theories for massive and massless fermions were studied in Refs.~\cite{Liu:2018xip,Liu:2020flb}).

\section{Wigner function up to linear order in $\hbar$ \label{sec:Wigner-function-at}}

We define the covariant Wigner function in an external classical electromagnetic field or U(1) gauge field as the Fourier transform of the two-point correlation function
\begin{equation}
W(x,p)\equiv\int\frac{d^{4}y}{(2\pi\hbar)^{4}}e^{-\frac{i}{\hbar}p\cdot y}U\left(x+\frac{y}{2},\,x-\frac{y}{2}\right)\left\langle \Omega\left|\hat{\bar{\psi}}\left(x+\frac{y}{2}\right)\otimes\hat{\psi}\left(x-\frac{y}{2}
\right)\right|\Omega\right\rangle,
\end{equation}
where the tensor product is defined as $[A\otimes B]_{ij}=A_jB_i$ with $i=1-4$ the Dirac index.
Here, $\left\langle \Omega\left|\hat{O}\right|\Omega\right\rangle $
represents the expectation value of the operator $\hat{O}$ on a given
quantum state $\left|\Omega\right\rangle $ and the gauge link $U\left(x+\frac{y}{2},\,x-\frac{y}{2}\right)$ is defined in a straight line
\begin{equation}
U\left(x+\frac{y}{2},\,x-\frac{y}{2}\right)\equiv \exp\left[-\frac{i}{\hbar}y^{\mu}\int_{-1/2}^{1/2}ds\,\mathbb{A}_{\mu}(x+s\,y)\right]\,
\end{equation}
with $\mathbb{A}_{\mu}$ being the U(1) gauge potential.
For a classical background field, the gauge link is purely a phase factor
instead of an operator. In order to clearly
display the physical meaning of the Wigner function,
it is advantageous to decompose $W(x,p)$ in terms of 16 independent generators
of the Clifford algebra, $\left\{ 1,\,i\gamma^{5},\,\gamma^{\mu},\,\gamma^{5}\gamma^{\mu},\,\sigma^{\mu\nu}\right\}$,
where $1$ is the $4\times4$ unit matrix,
$\gamma^{5}\equiv i\gamma^{0}\gamma^{1}\gamma^{2}\gamma^{3}$,
and $\sigma^{\mu\nu}\equiv\frac{i}{2}\left[\gamma^{\mu},\,\gamma^{\nu}\right]$,
\begin{equation}
W=\frac{1}{4}\left(\mathcal{F}+i\gamma^{5}\mathcal{P}
+\gamma^{\mu}\mathcal{V}_{\mu}+\gamma^{5}\gamma^{\mu}\mathcal{A}_{\mu}
+\frac{1}{2}\sigma^{\mu\nu}\mathcal{S}_{\mu\nu}\right).
\end{equation}
According to the parities and properties under Lorentz transformation,
the coefficients $\mathcal{F}$, $\mathcal{P}$, $\mathcal{V}_{\mu}$,
$\mathcal{A}_{\mu}$, and $\mathcal{S}_{\mu\nu}$ are called the
scalar, pseudo-scalar, vector, axial-vector, and tensor components
of the Wigner function. They are identified as the densities of some
physical quantities in phase space \cite{Vasak:1987um}. For example, $\mathcal{V}_{\mu}$
gives the fermion vector current density, $\mathcal{A}_{\mu}$ gives
the axial current or spin density, and $\mathcal{S}_{\mu\nu}$ gives the electric/magnetic dipole-moment
density.

Applying the Dirac equation to the Wigner function, one can derive the equation of motion for the Wigner function. The Wigner-function equation of motion has analytical solutions in some special cases,
such as in a constant electromagnetic field. In a general space-time
dependent electromagnetic field, the semi-classical expansion in
the reduced Planck's constant $\hbar$ provides a powerful method
to solve out the Wigner function order by order in $\hbar$.
The Wigner function at the lowest or zeroth order in $\hbar$ is
independent of spin, while at the first order in $\hbar$
the Wigner function contains the spin degrees of freedom.
In this paper, we only consider the Wigner function up to $\mathcal{O}(\hbar )$.
The most general solution up to this order
is given in Ref. \cite{Weickgenannt:2019dks},
\begin{eqnarray}
\mathcal{F} & = & m\left[V\delta(p^{2}-m^{2})-\frac{\hbar}{2}F^{\alpha\beta}
\Sigma_{\alpha\beta}\delta^{\prime}(p^{2}-m^{2})\right]+\mathcal{O}(\hbar^{2})\nonumber \,,\\
\mathcal{P} & = & \frac{\hbar}{4m}\epsilon^{\mu\nu\alpha\beta}\nabla_{\mu}\left[p_{\nu}\Sigma_{\alpha\beta}
\delta(p^{2}-m^{2})\right]+\mathcal{O}(\hbar^{2})\nonumber \,,\\
\mathcal{V}^{\mu} & = & p^{\mu}\left[V\delta(p^{2}-m^{2})-\frac{\hbar}{2}F^{\alpha\beta}
\Sigma_{\alpha\beta}\delta^{\prime}(p^{2}-m^{2})\right]+\frac{\hbar}{2}\nabla_{\nu}\left[\Sigma^{\mu\nu}
 \delta(p^{2}-m^{2})\right]+\mathcal{O}(\hbar^{2})\nonumber \,, \\
\mathcal{A}^{\mu} & = & n^{\mu}\delta(p^{2}-m^{2})+\hbar\tilde{F}^{\mu\nu}p_{\nu}V\delta^{\prime}(p^{2}-m^{2})
+\mathcal{O}(\hbar^{2})\nonumber \,, \\
\mathcal{S}^{\mu\nu} & = & m\left[\Sigma^{\mu\nu}\delta(p^{2}-m^{2})-\hbar F^{\mu\nu}V\delta^{\prime}(p^{2}-m^{2})\right]+\mathcal{O}(\hbar^{2}) \,,
\label{eq:Wigner function}
\end{eqnarray}
where the dipole-moment tensor satisfies the following constraint equation
\begin{equation}
\delta(p^{2}-m^{2})\left[p_{\nu}\Sigma^{\mu\nu}
-\frac{\hbar}{2}\nabla^{\mu}V\right]=\mathcal{O}(\hbar^{2}) \,,
\label{eq:constraint}
\end{equation}
with $\nabla^{\mu}\equiv\partial_{x}^{\mu}-F^{\mu\nu}\partial_{p\nu}$.
The spin polarization vector $n^{\mu}$ can be expressed by the dipole-moment tensor $\Sigma^{\mu\nu}$ as follows
\begin{equation}
n^{\mu}=-\frac{1}{2}\epsilon^{\mu\nu\alpha\beta}p_{\nu}\Sigma_{\alpha\beta}\, .
\label{eq:spin vector}
\end{equation}
The undetermined functions $V$ and $n^{\mu}$ satisfy the generalized
Boltzmann equation and the generalized Bargmann-Michel-Telegdi (BMT)
equation, respectively,
\begin{eqnarray}
0 & = & \delta(p^{2}-m^{2})\left[p\cdot\nabla V+\frac{\hbar}{4}(\partial_{x}^{\alpha}\tilde{F}^{\mu\nu})\partial_{p\alpha}\Sigma_{\mu\nu}\right]
-\delta^{\prime}(p^{2}-m^{2})\frac{\hbar}{2}F^{\mu\nu}p\cdot\nabla\Sigma_{\mu\nu}
 \nonumber \,, \\
0 & = & \delta(p^{2}-m^{2})\left[p\cdot\nabla n^{\mu}-F^{\mu\nu}n_{\nu}-\frac{\hbar}{2}p_{\nu}(\partial_{x}^{\alpha}\tilde{F}^{\mu\nu})
\partial_{p\alpha}V\right]+\delta^{\prime}(p^{2}-m^{2})\hbar\tilde{F}^{\mu\nu}p_{\nu}p\cdot\nabla V \, .
\label{eq:kinetic equations}
\end{eqnarray}
We note that the solution in Eqs. (\ref{eq:Wigner function}),
the constraint (\ref{eq:constraint}), and the kinetic equations (\ref{eq:kinetic equations})
are for massive fermions. In the massless case, the Wigner
function also has formal solution and the corresponding
chiral kinetic equations up to $\mathcal{O}(\hbar)$ can be derived.
In the rest part of this paper, we
will explicitly show that a smooth transition exists between the massive results
(\ref{eq:Wigner function}-\ref{eq:kinetic equations}) and corresponding
massless ones.

\section{Reference-frame dependence \label{sec:Reference-frame-dependence}}

The components of the Wigner function in Eqs. (\ref{eq:Wigner function})
are obviously Lorentz covariant. The quantities $V(x,p)$ and $\Sigma^{\mu\nu}(x,p)$
have clear physical meanings of fermion number (vector charge) density and dipole-moment tensor, respectively. Furthermore, in obtaining Eqs. (\ref{eq:Wigner function}), it is not necessary to introduce any additional reference-frame vector because one can always work in the co-moving frame of the massive particle.
However, this is not the case for massless fermions: the massless solutions in Refs. \cite{Hidaka:2016yjf,Huang:2018wdl,Gao:2018wmr,Liu:2018xip}
are inevitably reference-frame dependent.
Such a reference frame controls the way of decomposing a four-vector such as
$\mathcal{V}^{\mu}(x,p)$ and $\mathcal{A}^{\mu}(x,p)$ into one part parallel to $p^{\mu}$
and the other part perpendicular to $p^{\mu}$.
It also controls the way of decomposing an antisymmetric tensor such as $\Sigma^{\mu\nu}(x,p)$
into an `electric' part and a `magnetic' part.
Thus, in order to find a smooth transition to the massless fermions,
we need to first introduce a reference-frame to the Wigner-function solutions for massive fermions. We will achieve this by decomposing the dipole-moment tensor into the `electric' and a `magnetic' parts. In Appendix \ref{sec:alternative}, we discuss an alternative, but equivalent, way to introduce the reference frame.

It is well-known that the electromagnetic field tensor $F^{\mu\nu}$ can be
decomposed into the electric field $E^{\mu}=F^{\mu\nu}u_{\nu}$ and the magnetic field
$B^{\mu}=(1/2)\epsilon^{\mu\nu\alpha\beta}u_{\nu}F_{\alpha\beta}$ as
\begin{equation}
F^{\mu\nu}=E^{\mu}u^{\nu}-E^{\nu}u^{\mu}+\epsilon^{\mu\nu\alpha\beta}u_{\alpha}B_{\beta}\,,
\end{equation}
where $u^{\mu}$ is an arbitrary time-like vector which is normalized as
$u^{\mu}u_{\mu}=1$. The electric and magnetic field four-vectors are all
space-like, i.e. $u_{\mu}E^{\mu}=u_{\mu}B^{\mu}=0$. Thus in the
co-moving frame of $u^{\mu}$, $E^{\mu}$ and $B^{\mu}$ only have the spatial components
or they become three-vectors. Such a decomposition depends on the choice of $u^{\mu}$,
but $F^{\mu\nu}$ is independent of $u^{\mu}$. Similar to $F^{\mu\nu}$,
the dipole-moment tensor $\Sigma^{\mu\nu}$ can also be decomposed as
\begin{equation}
\Sigma^{\mu\nu}=\mathcal{E}^{\mu}u^{\nu}-\mathcal{E}^{\nu}u^{\mu}
-\epsilon^{\mu\nu\alpha\beta}u_{\alpha}\mathcal{M}_{\beta}\,,
\label{eq:Sigma decomposition}
\end{equation}
where $\mathcal{E}^{\mu}=\Sigma^{\mu\nu}u_{\nu}$ and $\mathcal{M}^{\mu}=-(1/2)\epsilon^{\mu\nu\alpha\beta}u_{\nu}\Sigma_{\alpha\beta}$.
Note that both $\mathcal{E}^{\mu}$ and $\mathcal{M}^{\mu}$ depend on
$u^{\mu}$, while $\Sigma^{\mu\nu}$ does not. In the co-moving frame of $u^{\mu}$,
we can identify $\mathcal{E}^{\mu}$ as the electric dipole moment and $\mathcal{M}^{\mu}$
as the magnetic dipole moment, respectively. Their physical meanings can be clearly
seen by contracting $F_{\mu\nu}$ and $\Sigma^{\mu\nu}$
\begin{equation}
-\frac{1}{2}F_{\mu\nu}\Sigma^{\mu\nu}=-\mathcal{E}_{\mu}E^{\mu}-\mathcal{M}_{\mu}B^{\mu} \,,
\end{equation}
which is the interaction energy of the dipole moments in the electromagnetic field.

Due to the constraint (\ref{eq:constraint}), the electric dipole moment $\mathcal{E}^{\mu}$ is not an independent degree of freedom. Substituting
Eq. (\ref{eq:Sigma decomposition})
into Eq. (\ref{eq:constraint}), up to $\mathcal{O}(\hbar)$, we obtain
\begin{equation}
\delta(p^{2}-m^{2})\left[(u\cdot p)\mathcal{E}^{\mu}-p_{\nu}\mathcal{E}^{\nu}u^{\mu}
-\epsilon^{\mu\nu\alpha\beta}p_{\nu}u_{\alpha}\mathcal{M}_{\beta}-\frac{\hbar}{2}\nabla^{\mu}V\right]=0 \,.
\label{eq:equation for Epsilon}
\end{equation}
Then contracting this equation with $u_{\mu}$ one obtains a constraint
for the electric dipole moment,
\begin{equation}
\delta(p^{2}-m^{2})\left(p_{\mu}\mathcal{E}^{\mu}+\frac{\hbar}{2}u_{\mu}\nabla^{\mu}V\right)=0 \,.
\label{eq:p dot E}
\end{equation}
Inserting Eq. (\ref{eq:p dot E}) back into Eq. (\ref{eq:equation for Epsilon}),
we find a general expression for the electric dipole moment
\begin{equation}
\mathcal{E}^{\mu}=\frac{\hbar}{2(u\cdot p)}(g^{\mu\nu}-u^{\mu}u^{\nu})\nabla_{\nu}V+\frac{1}{u\cdot p}\epsilon^{\mu\nu\alpha\beta}p_{\nu}u_{\alpha}\mathcal{M}_{\beta}+(p^{2}-m^{2})\mathcal{C}^{\mu} \,.
\label{eq:solution of E}
\end{equation}
where $\mathcal{C}^{\mu}$ is an arbitrary function
which should be non-singular for on-shell momentum with $p^{2}=m^{2}$.
We demand that $u_{\mu}\mathcal{C}^{\mu}=0$ because $u_{\mu}\mathcal{E}^{\mu}=0$.
Inserting Eq. (\ref{eq:solution of E}) back into Eq. (\ref{eq:Sigma decomposition}),
we find that $\Sigma^{\mu\nu}$ can be determined by $\mathcal{M}^{\mu}$, $V$
and $\mathcal{C}^{\mu}$,
\begin{equation}
\Sigma^{\mu\nu}=-\frac{1}{u\cdot p}\epsilon^{\mu\nu\alpha\beta}p_{\alpha}\mathcal{M}_{\beta}+\frac{\hbar}{2(u\cdot p)}\left(u^{\nu}\nabla^{\mu}-u^{\mu}\nabla^{\nu}\right)V
+(p^{2}-m^{2})\left(u^{\nu}\mathcal{C}^{\mu}-u^{\mu}\mathcal{C}^{\nu}\right)\,,
\label{eq:solution of Sigma}
\end{equation}
where we have used the Schouten identity
\begin{equation}
u^{\mu}\epsilon^{\nu\alpha\beta\gamma}+u^{\nu}\epsilon^{\alpha\beta\gamma\mu}
+u^{\alpha}\epsilon^{\beta\gamma\mu\nu}
+u^{\beta}\epsilon^{\gamma\mu\nu\alpha}+u^{\gamma}\epsilon^{\mu\nu\alpha\beta}=0 \,.
\label{eq:Schouten}
\end{equation}
In Eq. (\ref{eq:solution of Sigma}), we have an unspecified term which depends on $\mathcal{C}^{\mu}$.
However, since the forms of the solutions (\ref{eq:Wigner function}) do not change
under the transformation with arbitrary $\delta\Sigma^{\mu\nu}$ \cite{Weickgenannt:2019dks}
\begin{eqnarray}
\label{eq:shift}
\Sigma^{\mu\nu} & \rightarrow & \Sigma^{\prime\mu\nu}=\Sigma^{\mu\nu}+(p^{2}-m^{2})\delta\Sigma^{\mu\nu}\nonumber \,, \\
V & \rightarrow & V^{\prime}=V-\frac{\hbar}{2}F^{\mu\nu}\delta\Sigma_{\mu\nu} \,,
\end{eqnarray}
we can choose $\mathcal{C}^{\mu}=0$ without loss of generality.
When the electromagnetic field vanishes,
all components of the Wigner function are proportional to $\delta(p^{2}-m^{2})$
and thus the contribution from $\mathcal{C}^{\mu}$ is zero by its prefactor $(p^{2}-m^{2})$.

We further define the projection operators
\begin{equation}
\Delta^{\mu\nu}\equiv g^{\mu\nu}-u^{\mu}u^{\nu},\ \ \ \Xi^{\mu\nu}\equiv g^{\mu\nu}-u^{\mu}u^{\nu}+\frac{p^{\langle\mu\rangle}p^{\langle\nu\rangle}}{(u\cdot p)^{2}-p^{2}}\,, \label{projectionOps}
\end{equation}
where $p^{\langle\mu\rangle}\equiv\Delta^{\mu\nu}p_{\nu}$. The operator
$\Delta^{\mu\nu}$ projects a four-vector to the direction perpendicular
to $u^{\mu}$, while the operator $\Xi^{\mu\nu}$ projects it to the direction perpendicular
to both $u^{\mu}$ and $p^{\mu}$. Then we define the transverse magnetic dipole-moment
vector as follows,
\begin{equation}
\mathcal{M}_{\perp}^{\mu}\equiv\Xi^{\mu\nu}\mathcal{M}_{\nu}=\mathcal{M}^{\mu}-\frac{u\cdot p}{(u\cdot p)^{2}-p^{2}}p^{\langle\mu\rangle}A \,,
\end{equation}
where $A$ is defined as
\begin{equation}
A\equiv-\frac{\mathcal{M}\cdot p}{u\cdot p}
\end{equation}
and will be identified as the axial-charge density. Using $A$ and $\mathcal{M}_{\perp}^{\mu}$, the dipole-moment tensor $\Sigma^{\mu\nu}$
in Eq. (\ref{eq:solution of Sigma}) can be put into the form
\begin{equation}
\Sigma^{\mu\nu}=\frac{u\cdot p}{(u\cdot p)^{2}-p^{2}}\epsilon^{\mu\nu\alpha\beta}p_{\alpha}u_{\beta}A-\frac{1}{u\cdot p}\epsilon^{\mu\nu\alpha\beta}p_{\alpha}\mathcal{M}_{\perp\beta}
+\frac{\hbar}{2(u\cdot p)}\left(u^{\nu}\nabla^{\mu}-u^{\mu}\nabla^{\nu}\right)V \,.
\end{equation}

With the spin three-vector $\mathbf{n}$ in a particle's rest frame,
the particle is called longitudinally (transversely) polarized if
$\mathbf{n}$ is parallel (perpendicular) to $\mathbf{p}$.
Any other polarization state can be expressed as a superposition
of a longitudinal polarization state and a transverse one.
Generalizing such a three-dimensional decomposition to a four-dimensional one,
we can decompose the axial-vector $n^{\mu}$ as
\begin{equation}
n^{\mu}=(u\cdot p)\,u^{\mu}\frac{u\cdot n}{u\cdot p}
+n_{\parallel}p^{\langle\mu\rangle}+n_{\perp}^{\mu}\,,
\label{eq:massive spin polarization}
\end{equation}
where $n_{\perp}^{\mu}=\Xi^{\mu\nu}n_{\nu}$. In the co-moving frame of $u^{\mu}$,
$u\cdot n/u\cdot p$ is identified as the axial-charge density,
$n_{\parallel}$ is the longitudinal spin polarization and $n_{\perp}^{\mu}$
is the transverse spin polarization. Since $n^{\mu}$ satisfies the
constraint $p\cdot n =0$, $n_{\parallel}$ can be expressed by
\begin{equation}
n_{\parallel}=\frac{(u\cdot p)(u\cdot n)}{(u\cdot p)^{2}-p^{2}}\,.
\end{equation}
Inserting $\Sigma^{\mu\nu}$ into Eq. (\ref{eq:spin vector}),
we obtain another form of $n^{\mu}$,
\begin{eqnarray}
n^{\mu} & = & p^{\mu}A+\frac{p^{2}}{u\cdot p}\mathcal{M}^{\mu}+\frac{\hbar}{2(u\cdot p)}\epsilon^{\mu\nu\alpha\beta}p_{\nu}u_{\alpha}\nabla_{\beta}V\nonumber \\
 & = & p^{\mu}A+\frac{p^{2}}{(u\cdot p)^{2}-p^{2}}p^{\langle\mu\rangle}A+\frac{p^{2}}{u\cdot p}\mathcal{M}_{\perp}^{\mu}+\frac{\hbar}{2(u\cdot p)}\epsilon^{\mu\nu\alpha\beta}p_{\nu}u_{\alpha}\nabla_{\beta}V \,.
\label{eq:expression of nmu}
\end{eqnarray}
Making a comparison between Eq. (\ref{eq:massive spin polarization})
and Eq. (\ref{eq:expression of nmu}), we find
$A=u\cdot n/u\cdot p$ is the axial-charge density,
while the transverse polarization is
\begin{equation}
n_{\perp}^{\mu}=\frac{p^{2}}{u\cdot p}\mathcal{M}_{\perp}^{\mu}+\frac{\hbar}{2(u\cdot p)}\epsilon^{\mu\nu\alpha\beta}p_{\nu}u_{\alpha}\nabla_{\beta}V \,.
\label{eq:A and nparallel and nperp}
\end{equation}
In Sec. \ref{sec:Wigner-function-for}, we will show that the second
term of $n_{\perp}^{\mu}$ is the OAM of a wave packet, also known
as the side-jump term \cite{Chen:2014cla,Hidaka:2017auj,Huang:2018wdl,Gao:2018wmr}.
This term remains in the massless limit if we keep the wave-packet
description for massless particles. We will also prove that
$\mathcal{M}^{\mu}_\perp\propto 1/m$ in small-mass limit, which agrees with our knowledge about the magnetic moment
and is divergent at the zero mass limit.
However, $\mathcal{M}^{\mu}_\perp$ always comes with $p^{2}$ in $n_{\perp}^{\mu}$,
so its contribution, $p^{2}\mathcal{M}^{\mu}_\perp\propto m$, smoothly goes
towards zero for vanishing $m$.

\section{Wigner function for a wave packet\label{sec:Wigner-function-for}}

In this section, we introduce the wave-packet description of a single particle state into the Wigner function. For simplicity we neglect the external electromagnetic field.
The Wigner function is then given by
\begin{eqnarray}
W(x,p) & = & \int\frac{d^{4}q}{(2\pi\hbar)^{6}}\sum_{ss^{\prime}}\exp\left(\frac{i}{\hbar}q\cdot x\right)\delta\left(p^{2}+\frac{1}{4}q^{2}-m^{2}\right)\delta\left(p\cdot q\right)\sqrt{\left|p^{0}+\frac{1}{2}q^{0}\right|\left|p^{0}-\frac{1}{2}q^{0}\right|}\nonumber \\
 &  & \times\theta(p^{0})\bar{u}_{s}\left(\mathbf{p}+\frac{1}{2}\mathbf{q}\right)\otimes u_{s^{\prime}}\left(\mathbf{p}-\frac{1}{2}\mathbf{q}\right)\left\langle \Omega\left|\hat{a}_{\mathbf{p}+\frac{1}{2}\mathbf{q},s}^{\dagger}\hat{a}_{\mathbf{p}-\frac{1}{2}\mathbf{q},s^{\prime}}\right|\Omega\right\rangle. \label{eq:Wigner function-1}
\end{eqnarray}
Here we only keep the contribution from particles and neglect that from
anti-particles. The discussion for anti-particles can be similarly handled. The particle's spinors are denoted
as $u_s(\mathbf{p})$ and $\bar{u}_s(\mathbf{p})$ with $s$ denoting the spin state.
The state $\left|\Omega\right\rangle $ is assumed to take a wave-packet form
\begin{equation}
\left|\Omega\right\rangle =\left|\mathbf{p}_{0},s_{0},+\right\rangle \,_{\text{wp}}=\frac{1}{N}\int\frac{d^{3}\mathbf{p^{\prime}}}{(2\pi\hbar)^{3}}\exp\left[-\frac{(\mathbf{p}^{\prime}-\mathbf{p}_{0})^{2}}{4\sigma_{p}^{2}}+\frac{i}{\hbar}p^{\prime}\cdot x_{0}\right]a_{\mathbf{p}^{\prime},s_{0}}^{\dagger}\left|0\right\rangle ,\label{eq:wave packet state}
\end{equation}
where the normalization constant $N=\sqrt{\left[\sigma_{p}/\left(\hbar\sqrt{2\pi}\right)\right]^{3}}$
ensures the unit condition $\langle \Omega |\Omega \rangle =1$ and $p'^\mu=(E_{\mathbf{p}'}\equiv\sqrt{\mathbf{p'}^2+m^2},\mathbf{p}')$ is the on-shell four momentum.
Such a wave packet is the Gaussian type with the center momentum
$\mathbf{p}_{0}$ and center position $\mathbf{x}_{0}$ at time $t_{0}$.
The spin state of the wave packet is labelled as $s_{0}$
in a given spin quantization direction $\mathbf{n}_{0}$.
The momentum width of the wave packet is $\sigma_{p}$.
In general, we assume the wave packet to be narrow enough in momentum space so that
$\sigma_{p}\ll \left|\mathbf{p}_{0}\right|$. Thus we can assume that
the expectation value
$\left\langle \Omega\left|\hat{a}_{\mathbf{p}+\frac{1}{2}\mathbf{q},s}^{\dagger}\hat{a}_{\mathbf{p}-\frac{1}{2}
\mathbf{q},s^{\prime}}\right|\Omega\right\rangle $
vanish for a large $|\mathbf{q}|\gg\sigma_p$. One can then treat $\mathbf{q}$
as a small expansion variable in Eq. (\ref{eq:Wigner function-1})
except for $\exp\left(i q\cdot x/\hbar \right)$ and $\delta (p\cdot q)$.
Due to the existence of $\exp\left(i q\cdot x/\hbar \right)$, $\mathbf{q}$
can be replaced by $i\hbar\boldsymbol{\nabla}_{\mathbf{x}}$; in this sense,
the expansion in $\mathbf{q}$ is equivalent to the gradient expansion.
The leading and next-to-leading order terms in the Wigner function read
\begin{eqnarray}
W(x,p) & = & \frac{1}{(2\pi\hbar)^{3}}\delta(p^{2}-m^{2})\theta(p^{0})\sum_{ss^{\prime}} \left[\bar{u}_{s}(\mathbf{p})\otimes u_{s^{\prime}}(\mathbf{p})+\hbar\boldsymbol{\mathcal{U}}_{ss^{\prime}}\cdot i\boldsymbol{\nabla}_{\mathbf{x}}\right]f_{ss^{\prime}}(x,\mathbf{p}) ,
\label{eq:Wigner function leading two}
\end{eqnarray}
where the distribution function is defined as
\begin{equation}
f_{ss^{\prime}}(x,\mathbf{p})=\int\frac{d^{4}q}{(2\pi\hbar)^{3}}
\delta\left(q^{0}-\frac{\mathbf{p}\cdot\mathbf{q}}{E_{\mathbf{p}}}\right)\exp\left(\frac{i}{\hbar}q\cdot x\right)\left\langle \Omega\left|\hat{a}_{\mathbf{p}+\frac{1}{2}\mathbf{q},s}^{\dagger}
\hat{a}_{\mathbf{p}-\frac{1}{2}\mathbf{q},s^{\prime}}\right|\Omega\right\rangle .
\end{equation}
The three-vector $\boldsymbol{\mathcal{U}}_{ss^{\prime}}$ in Eq. (\ref{eq:Wigner function leading two})
contains momentum-derivatives of the spinors,
\begin{equation}
\boldsymbol{\mathcal{U}}_{ss^{\prime}}\equiv\frac{1}{2}\left\{ \left[\boldsymbol{\nabla}_{\mathbf{p}}\bar{u}_{s}(\mathbf{p})\right]\otimes u_{s^{\prime}}(\mathbf{p})-\bar{u}_{s}(\mathbf{p})\otimes\left[\boldsymbol{\nabla}_{\mathbf{p}}u_{s^{\prime}}(\mathbf{p})\right]\right\} ,
\end{equation}
which represents a Berry connection in Dirac space and its explicit form is calculated in Appendix \ref{sec:Berry-curvature}.
With Eq. (\ref{eq:wave packet state}), we can give
the analytical form of the distribution function as
\begin{equation}
f_{ss^{\prime}}(x,\mathbf{p})=(2\pi\hbar)^{3}V_{0}(x,\mathbf{p})\delta_{ss_{0}}\delta_{s^{\prime}s_{0}},
\label{f-ss-prime}
\end{equation}
where the Gaussian type distribution function $V_{0}(x,p)$ is given by
\begin{equation}
V_{0}(x,\mathbf{p})=\frac{8}{(2\pi\hbar)^{3}}\exp\left\{ -\frac{(\mathbf{p}-\mathbf{p}_{0})^{2}}{2\sigma_{p}^{2}}
-\frac{2\sigma_{p}^{2}}{\hbar^{2}}\left[(\mathbf{x}-\mathbf{x}_{0})
-\frac{\mathbf{p}}{E_{\mathbf{p}}}(t-t_{0})\right]^{2}\right\} .
\label{eq:wave packet}
\end{equation}
One can read in the above form of $V_{0}(x,p)$ the center momentum $\mathbf{p}_{0}$
and the center position $\mathbf{x}_{0}+(t-t_{0})\mathbf{p}/E_{\mathbf{p}}$.
Note that the center position moves with the velocity $\mathbf{p}/E_{\mathbf{p}}$,
reflecting the movement of the wave packet.
Since we have $\boldsymbol{\nabla}_{\mathbf{x}}V_{0}\propto\left[(\mathbf{x}-\mathbf{x}_{0})
-\mathbf{p}/E_{\mathbf{p}}(t-t_{0})\right]V_{0}$, we can identify $\left(\mathbf{p}\times\boldsymbol{\nabla}_{\mathbf{x}}\right)V_{0}\propto\left[\mathbf{p}\times(\mathbf{x}-\mathbf{x}_{0})\right]V_{0}$
as the OAM of the wave packet.
Inserting $f_{ss^{\prime}}(x,\mathbf{p})$ in (\ref{f-ss-prime}) into the Wigner function
in (\ref{eq:Wigner function leading two}), we derive the scalar component
of the Wigner function by taking the trace of $W$
\begin{equation}
\mathcal{F}=2\left[m-\frac{\hbar}{2(E_{\mathbf{p}}+m)}s_{0}\mathbf{n}_{0}\cdot(\mathbf{p}\times\boldsymbol{\nabla}_{\mathbf{x}})\right]V_{0}(x,\mathbf{p})\delta(p^{2}-m^{2})\theta(p^{0})\label{eq:scalar component of Wigner funcion}
\end{equation}
where $\mathbf{n}_{0}$ is the spin quantization direction, and
the second term on the right-hand-side is interpreted as
a correction from the spin-orbital coupling.
Equation (\ref{eq:scalar component of Wigner funcion}) is given in the lab
frame with $u^{\mu}=(1,0,0,0)$; a straightforward generalization to an arbitrary
frame gives
\begin{equation}
V(x,\mathbf{p})=2\left[1-\frac{\hbar}{2m(u\cdot p+m)}s_{0}\epsilon^{\mu\nu\alpha\beta}u_{\mu}n_{0\nu}p_{\alpha}\partial_{x\beta}\right]V_{0}(x,\mathbf{p}).
\end{equation}

Analogously, we can derive the axial-vector component of the Wigner function,
\begin{eqnarray}
\mathcal{A}^{0} & = &2\, s_{0}\,\mathbf{p}\cdot\mathbf{n}_{0}V_{0}(x,\mathbf{p})\delta(p^{2}-m^{2})\theta(p^{0}),\nonumber \\
\boldsymbol{\mathcal{A}} & = & 2 \left[s_{0}\,m\,\mathbf{n}_{0}+s_{0}\,
\frac{\mathbf{p}\cdot\mathbf{n}_{0}}{E_{\mathbf{p}}+m}\mathbf{p}-\frac{\hbar}{2(E_{\mathbf{p}}+m)}
\mathbf{p}\times\boldsymbol{\nabla}_{\mathbf{x}}\right]V_{0}(x,\mathbf{p})\delta(p^{2}-m^{2})\theta(p^{0}).
\end{eqnarray}
Comparing with Eq. (\ref{eq:massive spin polarization}) with the
reference frame taken as the lab frame $u^{\mu}=(1,0,0,0)$, we obtain
the axial-charge distribution $A$ and the transverse magnetic dipole moment as
\begin{eqnarray}
A(x,\mathbf{p}) & = &2\, s_{0}\,\frac{\mathbf{p}\cdot\mathbf{n}_{0}}{E_{\mathbf{p}}}V_{0}(x,\mathbf{p}),\nonumber \\
\boldsymbol{\mathcal{M}}_{\perp}(x,\mathbf{p}) & = & \frac{2}{m}\left[s_{0}\,E_{\mathbf{p}}\left(\mathbf{n}_{0}
-\frac{\mathbf{p}\cdot\mathbf{n}_{0}}{E_{\mathbf{p}}^{2}-m^{2}}\mathbf{p}\right)+\frac{\hbar}{2(E_{\mathbf{p}}+m)}\mathbf{p}
\times\boldsymbol{\nabla}_{\mathbf{x}}\right]V_{0}(x,\mathbf{p}).
\end{eqnarray}
Generalizing it to an arbitrary frame with a general $u^{\mu}$, we obtain %
\begin{eqnarray}
A & = & -2\, s_{0}\,\frac{p^{\alpha}\Delta_{\alpha\beta}n_{0}^{\beta}}{u\cdot p}V_{0}(x,p),\nonumber \\
\mathcal{M}_{\perp}^{\mu} & = & \frac{2}{m}\left[s_{0}(u\cdot p)\Xi ^\mu_\nu n^\nu_{0} +\frac{\hbar}{2(u\cdot p+m)}\epsilon^{\mu\nu\alpha\beta}u_{\nu}p_{\alpha}\partial_{x\beta}\right]V_{0}(x,p),
\label{eq:magnetic dipole moment}
\end{eqnarray}
where the projection operators $\Delta^{\mu\nu}$ and $\Xi^{\mu\nu}$ are defined in Eq. (\ref{projectionOps}). We see that the transverse magnetic dipole moment consists of two parts: one
is the particle's spin $s_{0}n_{0}^{\mu}$ as the intrinsic degrees of freedom
and the other is from the spatial derivative of the distribution
that can be identified as the OAM of the wave packet.

If there are many particles in the system, the calculation
of the Wigner function in the wave packet representation is straightforward:
we first calculate the each particle's Wigner function,
and then sum over all particles to obtain the total Wigner function. At equilibrium, this can also be achieved by replacing $\left|\Omega\right\rangle$ with the thermal state.

It can be easily verified that the small-mass behaviors of $V$, $A$, and $\mathcal{M}_{\perp}^{\mu}$ are
\begin{equation}
V=\mathcal{O}(m^{-1})+\mathcal{O}(1,\,m,\,\cdots),\ \ A=\mathcal{O}(1)+\mathcal{O}(m,\,m^{2},\,\cdots),\ \ \mathcal{M}_{\perp}^{\mu}=\mathcal{O}(m^{-1})+\mathcal{O}(1,\,m,\,\cdots) .
\end{equation}
In the next section, we will show that, in the Wigner function and
the kinetic equation, the divergent parts of $V$ and $\mathcal{M}_\perp^{\mu}$ at $m\rightarrow 0$
will either be cancelled or be suppressed by a factor $m^{2}$.

\section{Connection between massive and massless kinetic equations\label{sec:Mass-correction-to}}

For massless fermions, only the vector and axial-vector components
of the Wigner function are relevant to the kinetic equations. In this section
we focus on these components and show how to smoothly reproduce
the massless formula from the massive ones from Eq. (\ref{eq:Wigner function}).
We also show how to recover the CKT in the massless limit from the kinetic equations for massive fermions.

\subsection{Wigner function components}
Using the transverse part of the magnetic moment $\mathcal{M}_{\perp}^{\mu}$
and the axial-charge distribution $A$, the axial-vector component
of Wigner function is obtained by substituting Eq. (\ref{eq:expression of nmu})
into the axial-vector component in Eq. (\ref{eq:Wigner function}),
\begin{eqnarray}
\mathcal{A}^{\mu} & = & \left[(u\cdot p)u^{\mu}+\frac{(u\cdot p)^{2}}{(u\cdot p)^{2}-m^{2}}p^{\langle\mu\rangle}\right]A\delta(p^{2}-m^{2})\nonumber \\
 &  & +\frac{\hbar}{2(u\cdot p)}\epsilon^{\mu\nu\alpha\beta}p_{\nu}u_{\alpha}\left(\nabla_{\beta}V\right)
 \delta(p^{2}-m^{2})+\hbar\tilde{F}_{\mu\nu}p^{\nu}V\delta^{\prime}(p^{2}-m^{2})\nonumber \\
 &  & +\frac{m^{2}}{u\cdot p}\mathcal{M}_{\perp}^{\mu}\delta(p^{2}-m^{2}) .
 \label{eq:frame dependent Amu}
\end{eqnarray}
In the massless limit, we find that the first two lines agree with
the massless result in Refs. \cite{Hidaka:2016yjf,Huang:2018wdl,Gao:2018wmr,Liu:2018xip},
while the last line vanish because $\mathcal{M}_{\perp}^{\mu}\propto m^{-1}$.

On the other hand, inserting the dipole-moment tensor (\ref{eq:solution of Sigma})
into the vector component of the Wigner function in Eq. (\ref{eq:Wigner function}) gives
\begin{eqnarray}
\label{eq:vfinal}
\mathcal{V}^{\mu} & = & p^{\mu}V\delta(p^{2}-m^{2})+\hbar\tilde{F}^{\mu\nu}p_{\nu}A\delta^{\prime}(p^{2}-m^{2})+m^{2}\frac{\hbar}{u\cdot p}\tilde{F}^{\mu\nu}\mathcal{M}_{\nu}\delta^{\prime}(p^{2}-m^{2})\nonumber \\
 &  & -\frac{\hbar}{2}\delta(p^{2}-m^{2})\epsilon^{\mu\nu\alpha\beta}p_{\alpha}\nabla_{\nu}\left(\frac{1}{u\cdot p}\mathcal{M}_{\beta}\right),
\end{eqnarray}
where we have used the Schouten identity (\ref{eq:Schouten}).
It is not easy to see the small-mass-behaviour
of the last term, so we choose to rewrite it using the following relation
\begin{eqnarray}
\epsilon^{\mu\nu\alpha\beta}p_{\alpha}\nabla_{\nu}\left(\frac{1}{u\cdot p}\mathcal{M}_{\beta}\right) & = & -\frac{1}{u\cdot p}p^{\mu}u_{\rho}\epsilon^{\nu\alpha\beta\rho}p_{\alpha}\nabla_{\nu}\left(\frac{1}{u\cdot p}\mathcal{M}_{\beta}\right)-\frac{p^{2}}{u\cdot p}u_{\rho}\epsilon^{\beta\rho\mu\nu}\nabla_{\nu}\left(\frac{1}{u\cdot p}\mathcal{M}_{\beta}\right)\nonumber \\
 &  & -\frac{1}{u\cdot p}u_{\rho}p_{\alpha}\epsilon^{\alpha\beta\rho\mu}p^{\nu}\nabla_{\nu}\left(\frac{1}{u\cdot p}\mathcal{M}_{\beta}\right)-\frac{1}{u\cdot p}u_{\rho}p_{\alpha}p^{\beta}\epsilon^{\rho\mu\nu\alpha}\nabla_{\nu}\left(\frac{1}{u\cdot p}\mathcal{M}_{\beta}\right),
 \label{eq:temp equation}
\end{eqnarray}
which can be proved using the Schouten identity (\ref{eq:Schouten}).
Since the spin polarization $n^{\mu}$ satisfies the generalized BMT equation,
i.e. the second equation in Eq. (\ref{eq:kinetic equations}), we
obtain the following kinetic equation for the magnetic moment by replacing
$n^{\mu}$ with Eq. (\ref{eq:expression of nmu})
\begin{equation}
\delta(p^{2}-m^{2})p^{\nu}\nabla_{\nu}\left(\frac{1}{u\cdot p}\mathcal{M}^{\mu}\right)=\left[\frac{1}{u\cdot p}F^{\mu\nu}\mathcal{M}_{\nu}-\frac{1}{m^{2}}p^{\mu}\left(p^{\nu}\nabla_{\nu}A\right)\right]
\delta(p^{2}-m^{2})+\mathcal{O}(\hbar).
\end{equation}
Inserting the above relation and Eq. (\ref{eq:temp equation}) into Eq. (\ref{eq:vfinal}), we obtain %
\begin{eqnarray}
\mathcal{V}^{\mu} & = & p^{\mu}\tilde{V}\delta(p^{2}-m^{2})+\hbar\tilde{F}^{\mu\nu}p_{\nu}A\delta^{\prime}(p^{2}-m^{2})\nonumber \\
 &  & -\frac{\hbar}{2(u\cdot p)}\epsilon^{\mu\nu\alpha\beta}u_{\nu}p_{\alpha}\left(\nabla_{\beta}A\right)\delta(p^{2}-m^{2})\nonumber \\
 &  & +m^{2}\left\{ \frac{\hbar}{2(u\cdot p)}\epsilon^{\mu\nu\alpha\beta}u_{\nu}\left[\nabla_{\alpha}\left(\frac{1}{u\cdot p}\mathcal{M}_{\beta}\right)\right]\delta(p^{2}-m^{2})+\frac{\hbar}{u\cdot p}\tilde{F}^{\mu\nu}\mathcal{M}_{\nu}\delta^{\prime}(p^{2}-m^{2})\right\},
 \label{eq:frame-dependent Vmu2}
\end{eqnarray}
where we have redefined the distribution as
\begin{equation}
\tilde{V}\equiv V+\frac{\hbar}{2(u\cdot p)}\epsilon^{\alpha\beta\rho\sigma}u_{\alpha}p_{\beta}\nabla_{\rho}\left(\frac{1}{u\cdot p}\mathcal{M}_{\sigma}\right).
\end{equation}
In the massless limit, $\mathcal{V}^{\mu}$ in Eq. (\ref{eq:frame-dependent Vmu2})
smoothly reproduces the result in Refs. \cite{Hidaka:2016yjf,Huang:2018wdl,Gao:2018wmr,Liu:2018xip}. We
note that $\mathcal{M}_\perp^{\mu}\propto m^{-1}$ and $V\propto m^{-1}$ so that $\tilde{V}$ seems
to be divergent for small $m$. However, the divergent part of $V$
cancels exactly that of $\mathcal{M}^{\mu}$ leaving a finite $\tilde{V}$ in massless limit. In fact, taking a Gaussian wave packet as an example, as shown in Sec. \ref{sec:Wigner-function-for}, we obtain
\begin{equation}
\tilde{V}=\frac{1}{2}\left[1+\frac{\hbar}{2(u\cdot p)(u\cdot p+m)}s_{0}\epsilon^{\mu\nu\alpha\beta}u_{\mu}n_{0\nu}p_{\alpha}\partial_{x\beta}\right]V_{0}(x,\mathbf{p}),
\end{equation}
which is regular in $m\rightarrow0$ limit, where we have assumed a constant $u^{\mu}$ and vanishing electromagnetic field for simplicity and $V_{0}(x,\mathbf{p})$ is the distribution for the considered wave packet
given in Eq. (\ref{eq:wave packet}).

We note that the reference-frame four-vector $u^{\mu}$ can generally be a local vector in phase space, i.e. $u^{\mu}$ can be a function of $\{x^{\mu},p^{\mu}\}$.
Especially, if we take $u^{\mu}=p^{\mu}/m$, we would have
$\mathcal{M}^{\mu}=n^{\mu}/m$ and $A=0$, so $\mathcal{A}^{\mu}$ and
$\mathcal{V}^{\mu}$ in Eqs. (\ref{eq:frame dependent Amu}),(\ref{eq:frame-dependent Vmu2})
recover their forms in Eq. (\ref{eq:Wigner function}).

\subsection{Kinetic equations}

In the previous subsection, we have discussed the vector and axial-vector
components of the Wigner function. Since these components can smoothly recover
their forms in the massless case, it is natural that the corresponding kinetic
equations (\ref{eq:kinetic equations}) can reproduce the CKT when $m\rightarrow 0$. In this subsection, we explicitly show this.

In order to separate the kinetic equation for the axial-charge density
$A$, we first contract the generalized BMT equation (\ref{eq:kinetic equations})
with $u_{\mu}$ and then substitute $n^{\mu}$ with the expression
in Eq. (\ref{eq:expression of nmu}). After a long but straightforward
calculation and using the Schouten identity (\ref{eq:Schouten}),
we finally arrive at
\begin{eqnarray}
0 & = & \left(p^{\mu}\nabla_{\mu}A\right)\delta(p^{2}-m^{2})+\delta^{\prime}(p^{2}-m^{2})\left[\frac{\hbar}{u\cdot p}\tilde{F}^{\mu\nu}u_{\mu}p_{\nu}\left(p^{\alpha}\nabla_{\alpha}V\right)\right]\nonumber \\
 &  & +\delta(p^{2}-m^{2})\left[\frac{\hbar}{2}\epsilon^{\mu\nu\alpha\beta}p_{\nu}\left(\nabla_{\mu}\frac{u_{\alpha}}{u\cdot p}\right)\left(\nabla_{\beta}V\right)+\frac{\hbar}{2(u\cdot p)}p_{\mu}u_{\nu}(\partial_{x\alpha}\tilde{F}^{\mu\nu})(\partial_{p}^{\alpha}V)\right]\nonumber \\
 &  & -\delta(p^{2}-m^{2})\frac{m^{2}}{(u\cdot p)^{2}}\left[\left(p\cdot\nabla u_{\mu}-F_{\mu\nu}u^{\nu}\right)\mathcal{M}^{\mu}+\frac{\hbar}{2}\epsilon^{\mu\nu\alpha\beta}
 \left(\nabla_{\mu}V\right)u_{\nu}\nabla_{\alpha}u_{\beta}\right] .
 \label{eq:kinetic equation A}
\end{eqnarray}
Since $\mathcal{O}(\hbar^{2})$ terms are truncated throughout this
paper, one can replace $V$ in the above equation with $\tilde{V}$.
Meanwhile, the kinetic equation for the distribution $\tilde{V}$
can be derived from the generalized Boltzmann equation in Eq. (\ref{eq:kinetic equations})
by substituting the dipole-moment tensor $\Sigma^{\mu\nu}$ with the
expression (\ref{eq:solution of Sigma}). However, a simpler way
is to directly act $\nabla_{\mu}$ on the reference-frame dependent
$\mathcal{V}^{\mu}$ in Eq. (\ref{eq:frame-dependent Vmu2}). Properly
using the Schouten identity (\ref{eq:Schouten}) and the properties
of delta-functions
\begin{eqnarray}
x\delta^{\prime}(x) & = & -\delta(x),\nonumber \\
x\delta^{\prime\prime}(x) & = & -2\delta^{\prime}(x),
\end{eqnarray}
we derive the kinetic equation for $\tilde{V}$ as follows
\begin{eqnarray}
0 & = & \left(p^{\mu}\nabla_{\mu}\tilde{V}\right)\delta(p^{2}-m^{2})+\delta^{\prime}(p^{2}-m^{2})\left[\frac{\hbar}{u\cdot p}\tilde{F}^{\mu\nu}u_{\mu}p_{\nu}\left(p^{\alpha}\nabla_{\alpha}A\right)\right]\nonumber \\
 &  & +\delta(p^{2}-m^{2})\left[\frac{\hbar}{2}\epsilon^{\mu\nu\alpha\beta}p_{\nu}\left(\nabla_{\mu}\frac{u_{\alpha}}{u\cdot p}\right)\left(\nabla_{\beta}A\right)+\frac{\hbar}{2(u\cdot p)}p_{\mu}u_{\nu}(\partial_{x\alpha}\tilde{F}^{\mu\nu})(\partial_{p}^{\alpha}A)\right]\nonumber \\
 &  & +\delta(p^{2}-m^{2})m^{2}\left\{ \frac{\hbar}{2}\epsilon^{\mu\nu\alpha\beta}\left(\nabla_{\mu}\frac{u_{\nu}}{u\cdot p}\right)\left[\nabla_{\alpha}\left(\frac{1}{u\cdot p}\mathcal{M}_{\beta}\right)\right]-\frac{\hbar}{2(u\cdot p)}u_{\mu}(\partial_{x\alpha}\tilde{F}^{\mu\nu})\left[\partial_{p}^{\alpha}\left(\frac{1}{u\cdot p}\mathcal{M}_{\nu}\right)\right]\right\} \nonumber \\
 &  & +\delta^{\prime}(p^{2}-m^{2})m^{2}\frac{\hbar}{u\cdot p}\tilde{F}^{\mu\nu}u_{\mu}\left[p^{\alpha}\nabla_{\alpha}\left(\frac{1}{u\cdot p}\mathcal{M}_{\nu}\right)\right].\label{eq:kinetic equation V}
\end{eqnarray}
In the massless limit, Eqs. (\ref{eq:kinetic equation A}) and (\ref{eq:kinetic equation V})
agree exactly with the results in Refs. \cite{Hidaka:2016yjf,Huang:2018wdl,Gao:2018wmr,Liu:2018xip}.
Due to the chiral symmetry, equations for $A$ and $\tilde{V}$ have dual forms
in the massless case. However, the mass corrections in Eq.
(\ref{eq:kinetic equation A}) and (\ref{eq:kinetic equation V})
have very different forms.

As we have discussed in Sec. \ref{sec:Reference-frame-dependence},
$A$ and $\mathcal{M}_{\perp}^{\mu}$ are two independent variables
in describing spin polarization of massive fermions.
Now we have the kinetic equation for $A$ in (\ref{eq:kinetic equation A}).
The kinetic equation for $\mathcal{M}_{\perp}^{\mu}$
can be derived from the generalized BMT equation (\ref{eq:kinetic equations})
by substituting $n^{\mu}$ with the expression (\ref{eq:expression of nmu}).
The resulting equation is complicated which we would not show here.
This equation is, however, always accompanied with $m^2$ factor and becomes redundant in the massless limit. As a result, the kinetic equaions which describe four spin DOF of massive fermions, reduce to chiral kinetic equations describing two spin DOF of massless fermions.

\section{Summary }

We show how to smoothly connect the kinetic theories with
the spin degree of freedom for massive and massless fermions.
The Wigner-function components and the kinetic equations are
expressed in a reference-frame dependent form. The reference frame is introduced as the freedom
to decompose the dipole-moment tensor into an electric dipole-moment
vector and a magnetic dipole-moment one. Meanwhile, the spin polarization
is decomposed into an axial-charge distribution, a longitudinal polarization,
and a transverse polarization. Here a longitudinal (transverse) vector refers
to the one that is parallel (orthogonal) to the three-momentum in the
reference frame. The axial-charge distribution is obtained by projecting
the spin polarization onto the direction of the reference frame four-vector.
We find a straightforward relation between the longitudinal polarization
and the axial-charge distribution. Thus a minimum set of functions
for describing massive fermions is: the fermion distribution $V(x,p)$,
the axial-charge distribution $A(x,p)$, and the transverse part of
the magnetic dipole-moment $\mathcal{M}_{\perp}^{\mu}(x,p)$. By carefully
calculating these functions through the Wigner function in the wave packet representation,
we find their small-mass behaviors: $V\sim\mathcal{O}(m^{-1})$,
$\mathcal{M}_{\perp}^{\mu}\sim\mathcal{O}(m^{-1})$, and $A\sim\mathcal{O}(1)$.

With these small-mass behaviors, we can extract the mass corrections
in the vector and axial vector components of the Wigner function
as well as their corresponding kinetic equations. By turning off the mass corrections, we can smoothly recover their forms in the massless case.
Therefore the CKT can be obtained by a smooth transition from the kinetic theory for massive fermions with spin. We note that the collision terms are not included in this paper which are reserved for a future work. We expect that the side-jump effect can also arise naturally in the collision terms for massive fermions following the same line when taking the massless limit.

\section*{Acknowledgements}
The authors thank X.-Y. Guo, Y.-C. Liu, E. Speranza, and S. Pu for enlightening discussions. X.-L.S. and Q.W. are supported by the 973 program under Grant No.\ 2015CB856902 and by NSFC under Grant No.\ 11535012. X.-G.H is supported by NSFC under Grants No.\ 11535012 and No.\ 11675041.

After the completion of this work, we became aware of a related study \cite{Guo:2020gxy}.

\appendix

\section{An alternative way to introduce $u^\mu$ in $\mathcal{V}^{\mu}$ and $\mathcal{A}^{\mu}$\label{sec:alternative}}

In Sec.~\ref{sec:Reference-frame-dependence}, we introduce the reference frame by decomposing the dipole-moment tensor into the electric and magnetic components. In this Appendix, we adopt an alternative way to introduce the reference frame.

In Ref.~\cite{Liu:2020flb}, general forms of $\mathcal{V}^\mu$ and $\mathcal{A}^\mu$ up to $\mathcal{O}(\hbar)$ have been derived,
\begin{eqnarray}
\label{eq:b1}
\mathcal{V}^\mu&=&\delta(p^2-m^2)\left( p^\mu f +\frac{\hbar}{2u\cdot p}\epsilon^{\mu\nu\rho\sigma} u_\nu \nabla_\rho n_\sigma\right)+\hbar\tilde{F}^{\mu\nu}\left( n_\nu-u_\nu\frac{p\cdot n}{p\cdot u}\right)\delta'(p^2-m^2),\\
\label{eq:b2}\mathcal{A}^\mu&=&n^\mu\delta(p^2-m^2)+\hbar\tilde{F}^{\mu\nu}p_\nu f\delta'(p^2-m^2),
\end{eqnarray}
where the reference-frame vector $u^\mu$ is introduced when solving one of the equations of motion for the Wigner-function components (see, e.g. Refs.~\cite{Gao:2019znl,Weickgenannt:2019dks,Hattori:2019ahi,Liu:2020flb}): $(\hbar/2)(\nabla_\mu\mathcal{A}_\nu-\nabla_\nu\mathcal{A}_\mu)=\epsilon_{\mu\nu\rho\sigma}p^\rho\mathcal{V}^\sigma+\mathcal{O}(\hbar^2)$. Here $f$ is the vector charge density which is identical to $V$ in the main text at $\mathcal{O}(1)$ but can differ from $V$ at $\mathcal{O}(\hbar)$.  In fact, $\mathcal{V}^\mu$ and $\mathcal{A}^\mu$  in Eqs. (\ref{eq:b1}) and (\ref{eq:b2}) are equivalent to those in Eq. (\ref{eq:Wigner function}) because the solutions are invariant under transformations (\ref{eq:shift}) and  $n_\mu\rightarrow n_\mu+(p^2-m^2)\delta n_\mu, f\rightarrow f+\hbar\tilde{F}^{\mu\nu}u_\mu\delta n_\nu/u\cdot p$). Substituting Eqs. (\ref{eq:b1}) and (\ref{eq:b2}) into the relation (see, e.g. Refs.~\cite{Gao:2019znl,Weickgenannt:2019dks,Hattori:2019ahi,Liu:2020flb}) $m\mathcal{S}_{\mu\nu}=(\hbar/2)(\nabla_\mu\mathcal{V}_\nu-\nabla_\nu\mathcal{V}_\mu)-\epsilon_{\mu\nu\rho\sigma}p^\rho\mathcal{A}^\sigma+\mathcal{O}(\hbar^2)$, we obtain
\begin{eqnarray}
\label{eq:b3}
\mathcal{S}^{\mu\nu} & = & m\left[\Sigma^{\mu\nu}\delta(p^{2}-m^{2})-\hbar F^{\mu\nu}f\delta^{\prime}(p^{2}-m^{2})\right]+\mathcal{O}(\hbar^{2}).
\end{eqnarray}
In the above, we have defined the dipole-moment tensor as
\begin{equation}
\Sigma^{\mu\nu}=-\frac{1}{u\cdot p}\epsilon^{\mu\nu\alpha\beta}p_{\alpha}\mathcal{M}_{\beta}+\frac{\hbar}{2(u\cdot p)}\left(u^{\nu}\nabla^{\mu}-u^{\mu}\nabla^{\nu}\right)f\,,
\label{eq:b4}
\end{equation}
where
\begin{eqnarray}
\label{eq:b5}
\mathcal{M}^{\mu} &=& \frac{1}{m^2}\left[(n^\mu-p^\mu A)p\cdot u-\frac{\hbar}{2}\epsilon^{\mu\nu\alpha\beta}p_\nu u_\alpha\nabla_\beta f\right],\\
A&=&\frac{u\cdot n}{u\cdot p}=-\frac{\mathcal{M}\cdot p}{u\cdot p}.
\end{eqnarray}
Comparing to Eq. (\ref{eq:expression of nmu}) and noticing that $f=V +\mathcal{O}(\hbar)$, we realize that $\mathcal{M}^\mu$ we defined here is the magnetic dipole moment and the reference-frame vector $u^\mu$ is equivalent to the one introduced in the main text. From Eq. (\ref{eq:b5}) we can re-express $n^\mu$ in terms of $\mathcal{M}$ and $A$. By substituting $n^\mu$ into Eqs.~(\ref{eq:b1}) and (\ref{eq:b2}), after some algebra we recover Eqs. (\ref{eq:frame dependent Amu}) and Eq.~(\ref{eq:frame-dependent Vmu2}), but with
\begin{equation}
\tilde{V}\equiv f-\frac{\hbar}{(u\cdot p)^2}\tilde{F}^{\mu\nu}u_\mu\mathcal{M}_\nu.
\end{equation}

\section{Momentum-derivative of wavefunctions \label{sec:Berry-curvature}}

In this appendix, we will calculate the Berry connection $\boldsymbol{\mathcal{U}}_{ss^{\prime}}=\frac{1}{2}\left\{ \left[\boldsymbol{\nabla}_{\mathbf{p}}\bar{u}_{s}(\mathbf{p})\right]\otimes u_{s^{\prime}}(\mathbf{p})-\bar{u}_{s}(\mathbf{p})\otimes\left[\boldsymbol{\nabla}_{\mathbf{p}}u_{s^{\prime}}(\mathbf{p})\right]\right\} $.
Since we are considering massive fermions, we can express the wavefunction
$u_{s}(\mathbf{p})$ as a Lorentz boost of the wavefunction in the
rest frame $u_{s,\text{rf}}$,
\begin{equation}
u_{s}(\mathbf{p})=\Lambda_{\mathbf{p}}u_{s,\text{rf}}.
\end{equation}
In this way, all the momentum dependence is embedded in the transformation
matrix $\Lambda_{\mathbf{p}}$. The explicit form of $\Lambda_{\mathbf{p}}$
is well-known and can be found in many textbooks,
\begin{equation}
\Lambda_{\mathbf{p}}=\frac{1}{\sqrt{m}}\left(\begin{array}{cc}
\sqrt{p\cdot\sigma} & 0\\
0 & \sqrt{p\cdot\bar{\sigma}}
\end{array}\right),
\end{equation}
where $\sigma^{\mu}\equiv\left(1,\,\boldsymbol{\sigma}\right)$ and
$\bar{\sigma}^{\mu}\equiv\left(1,\,-\boldsymbol{\sigma}\right)$.
After complicated but straightforward calculations, we obtain the
following relation
\begin{equation}
\boldsymbol{\Delta}_{\mathbf{p}}\equiv\left(\boldsymbol{\nabla}_{\mathbf{p}}\Lambda_{\mathbf{p}}\right)\Lambda_{\mathbf{p}}^{-1}=\frac{1}{2m}\left[\gamma^{0}\boldsymbol{\gamma}-\frac{1}{E_{\mathbf{p}}(E_{\mathbf{p}}+m)}\mathbf{p}\left(\mathbf{p}\cdot\gamma^{0}\boldsymbol{\gamma}\right)\right]-\frac{i}{2m(E_{\mathbf{p}}+m)}\mathbf{p}\times\gamma^{5}\gamma^{0}\boldsymbol{\gamma}.
\end{equation}
Then the momentum derivatives of the wavefunctions $u_{s}(\mathbf{p})$
and $\bar{u}_{s}(\mathbf{p})$ are given by %
\begin{eqnarray}
\boldsymbol{\nabla}_{\mathbf{p}}u_{s}(\mathbf{p}) & = & \boldsymbol{\Delta}_{\mathbf{p}}u_{s}(\mathbf{p}),\nonumber \\
\boldsymbol{\nabla}_{\mathbf{p}}\bar{u}_{s}(\mathbf{p}) & = & -\bar{u}_{s}(\mathbf{p})\boldsymbol{\Delta}_{\mathbf{p}}.
\end{eqnarray}
The Berry connection now takes the
following form
\begin{equation}
\boldsymbol{\mathcal{U}}_{ss^{\prime}}=-\frac{1}{2}\left[\bar{u}_{s}(\mathbf{p})\boldsymbol{\Delta}_{\mathbf{p}}\otimes u_{s^{\prime}}(\mathbf{p})+\bar{u}_{s}(\mathbf{p})\otimes\boldsymbol{\Delta}_{\mathbf{p}}u_{s^{\prime}}(\mathbf{p})\right].
\end{equation}

\bibliographystyle{apsrev}
\bibliography{biblio_paper}

\begin{thebibliography}{72}
\expandafter\ifx\csname natexlab\endcsname\relax\def\natexlab#1{#1}\fi
\expandafter\ifx\csname bibnamefont\endcsname\relax
  \def\bibnamefont#1{#1}\fi
\expandafter\ifx\csname bibfnamefont\endcsname\relax
  \def\bibfnamefont#1{#1}\fi
\expandafter\ifx\csname citenamefont\endcsname\relax
  \def\citenamefont#1{#1}\fi
\expandafter\ifx\csname url\endcsname\relax
  \def\url#1{\texttt{#1}}\fi
\expandafter\ifx\csname urlprefix\endcsname\relax\def\urlprefix{URL }\fi
\providecommand{\bibinfo}[2]{#2}
\providecommand{\eprint}[2][]{\url{#2}}

\bibitem[{\citenamefont{Liang and Wang}(2005{\natexlab{a}})}]{Liang:2004ph}
\bibinfo{author}{\bibfnamefont{Z.-T.} \bibnamefont{Liang}} \bibnamefont{and}
  \bibinfo{author}{\bibfnamefont{X.-N.} \bibnamefont{Wang}},
  \bibinfo{journal}{Phys. Rev. Lett.} \textbf{\bibinfo{volume}{94}},
  \bibinfo{pages}{102301} (\bibinfo{year}{2005}{\natexlab{a}}),
  \bibinfo{note}{[Erratum: Phys. Rev. Lett.96,039901(2006)]},
  \eprint{nucl-th/0410079}.

\bibitem[{\citenamefont{Betz et~al.}(2007)\citenamefont{Betz, Gyulassy, and
  Torrieri}}]{Betz:2007kg}
\bibinfo{author}{\bibfnamefont{B.}~\bibnamefont{Betz}},
  \bibinfo{author}{\bibfnamefont{M.}~\bibnamefont{Gyulassy}}, \bibnamefont{and}
  \bibinfo{author}{\bibfnamefont{G.}~\bibnamefont{Torrieri}},
  \bibinfo{journal}{Phys. Rev.} \textbf{\bibinfo{volume}{C76}},
  \bibinfo{pages}{044901} (\bibinfo{year}{2007}), \eprint{0708.0035}.

\bibitem[{\citenamefont{Becattini et~al.}(2008)\citenamefont{Becattini,
  Piccinini, and Rizzo}}]{Becattini:2007sr}
\bibinfo{author}{\bibfnamefont{F.}~\bibnamefont{Becattini}},
  \bibinfo{author}{\bibfnamefont{F.}~\bibnamefont{Piccinini}},
  \bibnamefont{and} \bibinfo{author}{\bibfnamefont{J.}~\bibnamefont{Rizzo}},
  \bibinfo{journal}{Phys. Rev.} \textbf{\bibinfo{volume}{C77}},
  \bibinfo{pages}{024906} (\bibinfo{year}{2008}), \eprint{0711.1253}.

\bibitem[{\citenamefont{Skokov et~al.}(2009)\citenamefont{Skokov, Illarionov,
  and Toneev}}]{Skokov:2009qp}
\bibinfo{author}{\bibfnamefont{V.}~\bibnamefont{Skokov}},
  \bibinfo{author}{\bibfnamefont{A.~{\relax Yu}.} \bibnamefont{Illarionov}},
  \bibnamefont{and} \bibinfo{author}{\bibfnamefont{V.}~\bibnamefont{Toneev}},
  \bibinfo{journal}{Int. J. Mod. Phys.} \textbf{\bibinfo{volume}{A24}},
  \bibinfo{pages}{5925} (\bibinfo{year}{2009}), \eprint{0907.1396}.

\bibitem[{\citenamefont{Voronyuk et~al.}(2011)\citenamefont{Voronyuk, Toneev,
  Cassing, Bratkovskaya, Konchakovski, and Voloshin}}]{Voronyuk:2011jd}
\bibinfo{author}{\bibfnamefont{V.}~\bibnamefont{Voronyuk}},
  \bibinfo{author}{\bibfnamefont{V.~D.} \bibnamefont{Toneev}},
  \bibinfo{author}{\bibfnamefont{W.}~\bibnamefont{Cassing}},
  \bibinfo{author}{\bibfnamefont{E.~L.} \bibnamefont{Bratkovskaya}},
  \bibinfo{author}{\bibfnamefont{V.~P.} \bibnamefont{Konchakovski}},
  \bibnamefont{and} \bibinfo{author}{\bibfnamefont{S.~A.}
  \bibnamefont{Voloshin}}, \bibinfo{journal}{Phys. Rev.}
  \textbf{\bibinfo{volume}{C83}}, \bibinfo{pages}{054911}
  (\bibinfo{year}{2011}), \eprint{1103.4239}.

\bibitem[{\citenamefont{Deng and Huang}(2012)}]{Deng:2012pc}
\bibinfo{author}{\bibfnamefont{W.-T.} \bibnamefont{Deng}} \bibnamefont{and}
  \bibinfo{author}{\bibfnamefont{X.-G.} \bibnamefont{Huang}},
  \bibinfo{journal}{Phys. Rev.} \textbf{\bibinfo{volume}{C85}},
  \bibinfo{pages}{044907} (\bibinfo{year}{2012}), \eprint{1201.5108}.

\bibitem[{\citenamefont{McLerran and Skokov}(2014)}]{McLerran:2013hla}
\bibinfo{author}{\bibfnamefont{L.}~\bibnamefont{McLerran}} \bibnamefont{and}
  \bibinfo{author}{\bibfnamefont{V.}~\bibnamefont{Skokov}},
  \bibinfo{journal}{Nucl. Phys.} \textbf{\bibinfo{volume}{A929}},
  \bibinfo{pages}{184} (\bibinfo{year}{2014}), \eprint{1305.0774}.

\bibitem[{\citenamefont{Tuchin}(2013)}]{Tuchin:2013apa}
\bibinfo{author}{\bibfnamefont{K.}~\bibnamefont{Tuchin}},
  \bibinfo{journal}{Phys. Rev. C} \textbf{\bibinfo{volume}{88}},
  \bibinfo{pages}{024911} (\bibinfo{year}{2013}), \eprint{1305.5806}.

\bibitem[{\citenamefont{Deng and Huang}(2015)}]{Deng:2014uja}
\bibinfo{author}{\bibfnamefont{W.-T.} \bibnamefont{Deng}} \bibnamefont{and}
  \bibinfo{author}{\bibfnamefont{X.-G.} \bibnamefont{Huang}},
  \bibinfo{journal}{Phys. Lett. B} \textbf{\bibinfo{volume}{742}},
  \bibinfo{pages}{296} (\bibinfo{year}{2015}), \eprint{1411.2733}.

\bibitem[{\citenamefont{Li et~al.}(2016)\citenamefont{Li, Sheng, and
  Wang}}]{Li:2016tel}
\bibinfo{author}{\bibfnamefont{H.}~\bibnamefont{Li}},
  \bibinfo{author}{\bibfnamefont{X.-l.} \bibnamefont{Sheng}}, \bibnamefont{and}
  \bibinfo{author}{\bibfnamefont{Q.}~\bibnamefont{Wang}},
  \bibinfo{journal}{Phys. Rev.} \textbf{\bibinfo{volume}{C94}},
  \bibinfo{pages}{044903} (\bibinfo{year}{2016}), \eprint{1602.02223}.

\bibitem[{\citenamefont{Deng and Huang}(2016)}]{Deng:2016gyh}
\bibinfo{author}{\bibfnamefont{W.-T.} \bibnamefont{Deng}} \bibnamefont{and}
  \bibinfo{author}{\bibfnamefont{X.-G.} \bibnamefont{Huang}},
  \bibinfo{journal}{Phys. Rev. C} \textbf{\bibinfo{volume}{93}},
  \bibinfo{pages}{064907} (\bibinfo{year}{2016}), \eprint{1603.06117}.

\bibitem[{\citenamefont{Jiang et~al.}(2016)\citenamefont{Jiang, Lin, and
  Liao}}]{Jiang:2016woz}
\bibinfo{author}{\bibfnamefont{Y.}~\bibnamefont{Jiang}},
  \bibinfo{author}{\bibfnamefont{Z.-W.} \bibnamefont{Lin}}, \bibnamefont{and}
  \bibinfo{author}{\bibfnamefont{J.}~\bibnamefont{Liao}},
  \bibinfo{journal}{Phys. Rev. C} \textbf{\bibinfo{volume}{94}},
  \bibinfo{pages}{044910} (\bibinfo{year}{2016}), \bibinfo{note}{[Erratum:
  Phys.Rev.C 95, 049904 (2017)]}, \eprint{1602.06580}.

\bibitem[{\citenamefont{Pang et~al.}(2016)\citenamefont{Pang, Petersen, Wang,
  and Wang}}]{Pang:2016igs}
\bibinfo{author}{\bibfnamefont{L.-G.} \bibnamefont{Pang}},
  \bibinfo{author}{\bibfnamefont{H.}~\bibnamefont{Petersen}},
  \bibinfo{author}{\bibfnamefont{Q.}~\bibnamefont{Wang}}, \bibnamefont{and}
  \bibinfo{author}{\bibfnamefont{X.-N.} \bibnamefont{Wang}},
  \bibinfo{journal}{Phys. Rev. Lett.} \textbf{\bibinfo{volume}{117}},
  \bibinfo{pages}{192301} (\bibinfo{year}{2016}), \eprint{1605.04024}.

\bibitem[{\citenamefont{Deng et~al.}(2020)\citenamefont{Deng, Huang, Ma, and
  Zhang}}]{Deng:2020ygd}
\bibinfo{author}{\bibfnamefont{X.-G.} \bibnamefont{Deng}},
  \bibinfo{author}{\bibfnamefont{X.-G.} \bibnamefont{Huang}},
  \bibinfo{author}{\bibfnamefont{Y.-G.} \bibnamefont{Ma}}, \bibnamefont{and}
  \bibinfo{author}{\bibfnamefont{S.}~\bibnamefont{Zhang}}
  (\bibinfo{year}{2020}), \eprint{2001.01371}.

\bibitem[{\citenamefont{Voloshin}(2004)}]{Voloshin:2004ha}
\bibinfo{author}{\bibfnamefont{S.~A.} \bibnamefont{Voloshin}}
  (\bibinfo{year}{2004}), \eprint{nucl-th/0410089}.

\bibitem[{\citenamefont{Fang et~al.}(2016)\citenamefont{Fang, Pang, Wang, and
  Wang}}]{Fang:2016vpj}
\bibinfo{author}{\bibfnamefont{R.-h.} \bibnamefont{Fang}},
  \bibinfo{author}{\bibfnamefont{L.-g.} \bibnamefont{Pang}},
  \bibinfo{author}{\bibfnamefont{Q.}~\bibnamefont{Wang}}, \bibnamefont{and}
  \bibinfo{author}{\bibfnamefont{X.-n.} \bibnamefont{Wang}},
  \bibinfo{journal}{Phys. Rev.} \textbf{\bibinfo{volume}{C94}},
  \bibinfo{pages}{024904} (\bibinfo{year}{2016}), \eprint{1604.04036}.

\bibitem[{\citenamefont{Gao et~al.}(2008)\citenamefont{Gao, Chen, Deng, Liang,
  Wang, and Wang}}]{Gao:2007bc}
\bibinfo{author}{\bibfnamefont{J.-H.} \bibnamefont{Gao}},
  \bibinfo{author}{\bibfnamefont{S.-W.} \bibnamefont{Chen}},
  \bibinfo{author}{\bibfnamefont{W.-t.} \bibnamefont{Deng}},
  \bibinfo{author}{\bibfnamefont{Z.-T.} \bibnamefont{Liang}},
  \bibinfo{author}{\bibfnamefont{Q.}~\bibnamefont{Wang}}, \bibnamefont{and}
  \bibinfo{author}{\bibfnamefont{X.-N.} \bibnamefont{Wang}},
  \bibinfo{journal}{Phys. Rev.} \textbf{\bibinfo{volume}{C77}},
  \bibinfo{pages}{044902} (\bibinfo{year}{2008}), \eprint{0710.2943}.

\bibitem[{\citenamefont{Huang et~al.}(2011)\citenamefont{Huang, Huovinen, and
  Wang}}]{Huang:2011ru}
\bibinfo{author}{\bibfnamefont{X.-G.} \bibnamefont{Huang}},
  \bibinfo{author}{\bibfnamefont{P.}~\bibnamefont{Huovinen}}, \bibnamefont{and}
  \bibinfo{author}{\bibfnamefont{X.-N.} \bibnamefont{Wang}},
  \bibinfo{journal}{Phys. Rev. C} \textbf{\bibinfo{volume}{84}},
  \bibinfo{pages}{054910} (\bibinfo{year}{2011}), \eprint{1108.5649}.

\bibitem[{\citenamefont{Becattini et~al.}(2013)\citenamefont{Becattini,
  Chandra, Del~Zanna, and Grossi}}]{Becattini:2013fla}
\bibinfo{author}{\bibfnamefont{F.}~\bibnamefont{Becattini}},
  \bibinfo{author}{\bibfnamefont{V.}~\bibnamefont{Chandra}},
  \bibinfo{author}{\bibfnamefont{L.}~\bibnamefont{Del~Zanna}},
  \bibnamefont{and} \bibinfo{author}{\bibfnamefont{E.}~\bibnamefont{Grossi}},
  \bibinfo{journal}{Annals Phys.} \textbf{\bibinfo{volume}{338}},
  \bibinfo{pages}{32} (\bibinfo{year}{2013}), \eprint{1303.3431}.

\bibitem[{\citenamefont{Zhang et~al.}(2019)\citenamefont{Zhang, Fang, Wang, and
  Wang}}]{Zhang:2019xya}
\bibinfo{author}{\bibfnamefont{J.-j.} \bibnamefont{Zhang}},
  \bibinfo{author}{\bibfnamefont{R.-h.} \bibnamefont{Fang}},
  \bibinfo{author}{\bibfnamefont{Q.}~\bibnamefont{Wang}}, \bibnamefont{and}
  \bibinfo{author}{\bibfnamefont{X.-N.} \bibnamefont{Wang}},
  \bibinfo{journal}{Phys. Rev.} \textbf{\bibinfo{volume}{C100}},
  \bibinfo{pages}{064904} (\bibinfo{year}{2019}), \eprint{1904.09152}.

\bibitem[{\citenamefont{Adamczyk et~al.}(2017)}]{STAR:2017ckg}
\bibinfo{author}{\bibfnamefont{L.}~\bibnamefont{Adamczyk}} \bibnamefont{et~al.}
  (\bibinfo{collaboration}{STAR}), \bibinfo{journal}{Nature}
  \textbf{\bibinfo{volume}{548}}, \bibinfo{pages}{62} (\bibinfo{year}{2017}),
  \eprint{1701.06657}.

\bibitem[{\citenamefont{Adam et~al.}(2018)}]{Adam:2018ivw}
\bibinfo{author}{\bibfnamefont{J.}~\bibnamefont{Adam}} \bibnamefont{et~al.}
  (\bibinfo{collaboration}{STAR}), \bibinfo{journal}{Phys. Rev.}
  \textbf{\bibinfo{volume}{C98}}, \bibinfo{pages}{014910}
  (\bibinfo{year}{2018}), \eprint{1805.04400}.

\bibitem[{\citenamefont{Wang}(2017)}]{Wang:2017jpl}
\bibinfo{author}{\bibfnamefont{Q.}~\bibnamefont{Wang}}, \bibinfo{journal}{Nucl.
  Phys.} \textbf{\bibinfo{volume}{A967}}, \bibinfo{pages}{225}
  (\bibinfo{year}{2017}), \eprint{1704.04022}.

\bibitem[{\citenamefont{Huang}(2020)}]{Huang:2020xyr}
\bibinfo{author}{\bibfnamefont{X.-G.} \bibnamefont{Huang}}
  (\bibinfo{year}{2020}), \eprint{2002.07549}.

\bibitem[{\citenamefont{Becattini and Lisa}(2020)}]{Becattini:2020ngo}
\bibinfo{author}{\bibfnamefont{F.}~\bibnamefont{Becattini}} \bibnamefont{and}
  \bibinfo{author}{\bibfnamefont{M.~A.} \bibnamefont{Lisa}}
  (\bibinfo{year}{2020}), \eprint{2003.03640}.

\bibitem[{\citenamefont{Vilenkin}(1980)}]{Vilenkin:1980fu}
\bibinfo{author}{\bibfnamefont{A.}~\bibnamefont{Vilenkin}},
  \bibinfo{journal}{Phys. Rev.} \textbf{\bibinfo{volume}{D22}},
  \bibinfo{pages}{3080} (\bibinfo{year}{1980}).

\bibitem[{\citenamefont{Kharzeev et~al.}(2008)\citenamefont{Kharzeev, McLerran,
  and Warringa}}]{Kharzeev:2007jp}
\bibinfo{author}{\bibfnamefont{D.~E.} \bibnamefont{Kharzeev}},
  \bibinfo{author}{\bibfnamefont{L.~D.} \bibnamefont{McLerran}},
  \bibnamefont{and} \bibinfo{author}{\bibfnamefont{H.~J.}
  \bibnamefont{Warringa}}, \bibinfo{journal}{Nucl. Phys.}
  \textbf{\bibinfo{volume}{A803}}, \bibinfo{pages}{227} (\bibinfo{year}{2008}),
  \eprint{0711.0950}.

\bibitem[{\citenamefont{Fukushima et~al.}(2008)\citenamefont{Fukushima,
  Kharzeev, and Warringa}}]{Fukushima:2008xe}
\bibinfo{author}{\bibfnamefont{K.}~\bibnamefont{Fukushima}},
  \bibinfo{author}{\bibfnamefont{D.~E.} \bibnamefont{Kharzeev}},
  \bibnamefont{and} \bibinfo{author}{\bibfnamefont{H.~J.}
  \bibnamefont{Warringa}}, \bibinfo{journal}{Phys. Rev.}
  \textbf{\bibinfo{volume}{D78}}, \bibinfo{pages}{074033}
  (\bibinfo{year}{2008}), \eprint{0808.3382}.

\bibitem[{\citenamefont{Kharzeev et~al.}(2016)\citenamefont{Kharzeev, Liao,
  Voloshin, and Wang}}]{Kharzeev:2015znc}
\bibinfo{author}{\bibfnamefont{D.~E.} \bibnamefont{Kharzeev}},
  \bibinfo{author}{\bibfnamefont{J.}~\bibnamefont{Liao}},
  \bibinfo{author}{\bibfnamefont{S.~A.} \bibnamefont{Voloshin}},
  \bibnamefont{and} \bibinfo{author}{\bibfnamefont{G.}~\bibnamefont{Wang}},
  \bibinfo{journal}{Prog. Part. Nucl. Phys.} \textbf{\bibinfo{volume}{88}},
  \bibinfo{pages}{1} (\bibinfo{year}{2016}), \eprint{1511.04050}.

\bibitem[{\citenamefont{Huang}(2016)}]{Huang:2015oca}
\bibinfo{author}{\bibfnamefont{X.-G.} \bibnamefont{Huang}},
  \bibinfo{journal}{Rept. Prog. Phys.} \textbf{\bibinfo{volume}{79}},
  \bibinfo{pages}{076302} (\bibinfo{year}{2016}), \eprint{1509.04073}.

\bibitem[{\citenamefont{Hattori and Huang}(2017)}]{Hattori:2016emy}
\bibinfo{author}{\bibfnamefont{K.}~\bibnamefont{Hattori}} \bibnamefont{and}
  \bibinfo{author}{\bibfnamefont{X.-G.} \bibnamefont{Huang}},
  \bibinfo{journal}{Nucl. Sci. Tech.} \textbf{\bibinfo{volume}{28}},
  \bibinfo{pages}{26} (\bibinfo{year}{2017}), \eprint{1609.00747}.

\bibitem[{\citenamefont{Zhao and Wang}(2019)}]{Zhao:2019hta}
\bibinfo{author}{\bibfnamefont{J.}~\bibnamefont{Zhao}} \bibnamefont{and}
  \bibinfo{author}{\bibfnamefont{F.}~\bibnamefont{Wang}},
  \bibinfo{journal}{Prog. Part. Nucl. Phys.} \textbf{\bibinfo{volume}{107}},
  \bibinfo{pages}{200} (\bibinfo{year}{2019}), \eprint{1906.11413}.

\bibitem[{\citenamefont{Li and Wang}(2020)}]{Li:2020dwr}
\bibinfo{author}{\bibfnamefont{W.}~\bibnamefont{Li}} \bibnamefont{and}
  \bibinfo{author}{\bibfnamefont{G.}~\bibnamefont{Wang}}
  (\bibinfo{year}{2020}), \eprint{2002.10397}.

\bibitem[{\citenamefont{Liu and Huang}(2020)}]{Liu:2020ymh}
\bibinfo{author}{\bibfnamefont{Y.-C.} \bibnamefont{Liu}} \bibnamefont{and}
  \bibinfo{author}{\bibfnamefont{X.-G.} \bibnamefont{Huang}}
  (\bibinfo{year}{2020}), \eprint{2003.12482}.

\bibitem[{\citenamefont{Son and Yamamoto}(2012)}]{Son:2012wh}
\bibinfo{author}{\bibfnamefont{D.~T.} \bibnamefont{Son}} \bibnamefont{and}
  \bibinfo{author}{\bibfnamefont{N.}~\bibnamefont{Yamamoto}},
  \bibinfo{journal}{Phys. Rev. Lett.} \textbf{\bibinfo{volume}{109}},
  \bibinfo{pages}{181602} (\bibinfo{year}{2012}), \eprint{1203.2697}.

\bibitem[{\citenamefont{Son and Yamamoto}(2013)}]{Son:2012zy}
\bibinfo{author}{\bibfnamefont{D.~T.} \bibnamefont{Son}} \bibnamefont{and}
  \bibinfo{author}{\bibfnamefont{N.}~\bibnamefont{Yamamoto}},
  \bibinfo{journal}{Phys. Rev. D} \textbf{\bibinfo{volume}{87}},
  \bibinfo{pages}{085016} (\bibinfo{year}{2013}), \eprint{1210.8158}.

\bibitem[{\citenamefont{Stephanov and Yin}(2012)}]{Stephanov:2012ki}
\bibinfo{author}{\bibfnamefont{M.~A.} \bibnamefont{Stephanov}}
  \bibnamefont{and} \bibinfo{author}{\bibfnamefont{Y.}~\bibnamefont{Yin}},
  \bibinfo{journal}{Phys. Rev. Lett.} \textbf{\bibinfo{volume}{109}},
  \bibinfo{pages}{162001} (\bibinfo{year}{2012}), \eprint{1207.0747}.

\bibitem[{\citenamefont{Gao et~al.}(2012)\citenamefont{Gao, Liang, Pu, Wang,
  and Wang}}]{Gao:2012ix}
\bibinfo{author}{\bibfnamefont{J.-H.} \bibnamefont{Gao}},
  \bibinfo{author}{\bibfnamefont{Z.-T.} \bibnamefont{Liang}},
  \bibinfo{author}{\bibfnamefont{S.}~\bibnamefont{Pu}},
  \bibinfo{author}{\bibfnamefont{Q.}~\bibnamefont{Wang}}, \bibnamefont{and}
  \bibinfo{author}{\bibfnamefont{X.-N.} \bibnamefont{Wang}},
  \bibinfo{journal}{Phys. Rev. Lett.} \textbf{\bibinfo{volume}{109}},
  \bibinfo{pages}{232301} (\bibinfo{year}{2012}), \eprint{1203.0725}.

\bibitem[{\citenamefont{Chen et~al.}(2013)\citenamefont{Chen, Pu, Wang, and
  Wang}}]{Chen:2012ca}
\bibinfo{author}{\bibfnamefont{J.-W.} \bibnamefont{Chen}},
  \bibinfo{author}{\bibfnamefont{S.}~\bibnamefont{Pu}},
  \bibinfo{author}{\bibfnamefont{Q.}~\bibnamefont{Wang}}, \bibnamefont{and}
  \bibinfo{author}{\bibfnamefont{X.-N.} \bibnamefont{Wang}},
  \bibinfo{journal}{Phys. Rev. Lett.} \textbf{\bibinfo{volume}{110}},
  \bibinfo{pages}{262301} (\bibinfo{year}{2013}), \eprint{1210.8312}.

\bibitem[{\citenamefont{Hidaka et~al.}(2017)\citenamefont{Hidaka, Pu, and
  Yang}}]{Hidaka:2016yjf}
\bibinfo{author}{\bibfnamefont{Y.}~\bibnamefont{Hidaka}},
  \bibinfo{author}{\bibfnamefont{S.}~\bibnamefont{Pu}}, \bibnamefont{and}
  \bibinfo{author}{\bibfnamefont{D.-L.} \bibnamefont{Yang}},
  \bibinfo{journal}{Phys. Rev.} \textbf{\bibinfo{volume}{D95}},
  \bibinfo{pages}{091901} (\bibinfo{year}{2017}), \eprint{1612.04630}.

\bibitem[{\citenamefont{Huang et~al.}(2018)\citenamefont{Huang, Shi, Jiang,
  Liao, and Zhuang}}]{Huang:2018wdl}
\bibinfo{author}{\bibfnamefont{A.}~\bibnamefont{Huang}},
  \bibinfo{author}{\bibfnamefont{S.}~\bibnamefont{Shi}},
  \bibinfo{author}{\bibfnamefont{Y.}~\bibnamefont{Jiang}},
  \bibinfo{author}{\bibfnamefont{J.}~\bibnamefont{Liao}}, \bibnamefont{and}
  \bibinfo{author}{\bibfnamefont{P.}~\bibnamefont{Zhuang}},
  \bibinfo{journal}{Phys. Rev.} \textbf{\bibinfo{volume}{D98}},
  \bibinfo{pages}{036010} (\bibinfo{year}{2018}), \eprint{1801.03640}.

\bibitem[{\citenamefont{Gao et~al.}(2018{\natexlab{a}})\citenamefont{Gao,
  Liang, Wang, and Wang}}]{Gao:2018wmr}
\bibinfo{author}{\bibfnamefont{J.-H.} \bibnamefont{Gao}},
  \bibinfo{author}{\bibfnamefont{Z.-T.} \bibnamefont{Liang}},
  \bibinfo{author}{\bibfnamefont{Q.}~\bibnamefont{Wang}}, \bibnamefont{and}
  \bibinfo{author}{\bibfnamefont{X.-N.} \bibnamefont{Wang}},
  \bibinfo{journal}{Phys. Rev.} \textbf{\bibinfo{volume}{D98}},
  \bibinfo{pages}{036019} (\bibinfo{year}{2018}{\natexlab{a}}),
  \eprint{1802.06216}.

\bibitem[{\citenamefont{Liu et~al.}(2019{\natexlab{a}})\citenamefont{Liu, Gao,
  Mameda, and Huang}}]{Liu:2018xip}
\bibinfo{author}{\bibfnamefont{Y.-C.} \bibnamefont{Liu}},
  \bibinfo{author}{\bibfnamefont{L.-L.} \bibnamefont{Gao}},
  \bibinfo{author}{\bibfnamefont{K.}~\bibnamefont{Mameda}}, \bibnamefont{and}
  \bibinfo{author}{\bibfnamefont{X.-G.} \bibnamefont{Huang}},
  \bibinfo{journal}{Phys. Rev. D} \textbf{\bibinfo{volume}{99}},
  \bibinfo{pages}{085014} (\bibinfo{year}{2019}{\natexlab{a}}),
  \eprint{1812.10127}.

\bibitem[{\citenamefont{Hidaka et~al.}(2018)\citenamefont{Hidaka, Pu, and
  Yang}}]{Hidaka:2017auj}
\bibinfo{author}{\bibfnamefont{Y.}~\bibnamefont{Hidaka}},
  \bibinfo{author}{\bibfnamefont{S.}~\bibnamefont{Pu}}, \bibnamefont{and}
  \bibinfo{author}{\bibfnamefont{D.-L.} \bibnamefont{Yang}},
  \bibinfo{journal}{Phys. Rev.} \textbf{\bibinfo{volume}{D97}},
  \bibinfo{pages}{016004} (\bibinfo{year}{2018}), \eprint{1710.00278}.

\bibitem[{\citenamefont{Gao et~al.}(2017)\citenamefont{Gao, Pu, and
  Wang}}]{Gao:2017gfq}
\bibinfo{author}{\bibfnamefont{J.-h.} \bibnamefont{Gao}},
  \bibinfo{author}{\bibfnamefont{S.}~\bibnamefont{Pu}}, \bibnamefont{and}
  \bibinfo{author}{\bibfnamefont{Q.}~\bibnamefont{Wang}},
  \bibinfo{journal}{Phys. Rev.} \textbf{\bibinfo{volume}{D96}},
  \bibinfo{pages}{016002} (\bibinfo{year}{2017}), \eprint{1704.00244}.

\bibitem[{\citenamefont{Carignano et~al.}(2018)\citenamefont{Carignano, Manuel,
  and Torres-Rincon}}]{Carignano:2018gqt}
\bibinfo{author}{\bibfnamefont{S.}~\bibnamefont{Carignano}},
  \bibinfo{author}{\bibfnamefont{C.}~\bibnamefont{Manuel}}, \bibnamefont{and}
  \bibinfo{author}{\bibfnamefont{J.~M.} \bibnamefont{Torres-Rincon}},
  \bibinfo{journal}{Phys. Rev. D} \textbf{\bibinfo{volume}{98}},
  \bibinfo{pages}{076005} (\bibinfo{year}{2018}), \eprint{1806.01684}.

\bibitem[{\citenamefont{Lin and Shukla}(2019)}]{Lin:2019ytz}
\bibinfo{author}{\bibfnamefont{S.}~\bibnamefont{Lin}} \bibnamefont{and}
  \bibinfo{author}{\bibfnamefont{A.}~\bibnamefont{Shukla}},
  \bibinfo{journal}{JHEP} \textbf{\bibinfo{volume}{06}}, \bibinfo{pages}{060}
  (\bibinfo{year}{2019}), \eprint{1901.01528}.

\bibitem[{\citenamefont{Huang and Sadofyev}(2019)}]{Huang:2018aly}
\bibinfo{author}{\bibfnamefont{X.-G.} \bibnamefont{Huang}} \bibnamefont{and}
  \bibinfo{author}{\bibfnamefont{A.~V.} \bibnamefont{Sadofyev}},
  \bibinfo{journal}{JHEP} \textbf{\bibinfo{volume}{03}}, \bibinfo{pages}{084}
  (\bibinfo{year}{2019}), \eprint{1805.08779}.

\bibitem[{\citenamefont{Chen et~al.}(2014)\citenamefont{Chen, Son, Stephanov,
  Yee, and Yin}}]{Chen:2014cla}
\bibinfo{author}{\bibfnamefont{J.-Y.} \bibnamefont{Chen}},
  \bibinfo{author}{\bibfnamefont{D.~T.} \bibnamefont{Son}},
  \bibinfo{author}{\bibfnamefont{M.~A.} \bibnamefont{Stephanov}},
  \bibinfo{author}{\bibfnamefont{H.-U.} \bibnamefont{Yee}}, \bibnamefont{and}
  \bibinfo{author}{\bibfnamefont{Y.}~\bibnamefont{Yin}},
  \bibinfo{journal}{Phys. Rev. Lett.} \textbf{\bibinfo{volume}{113}},
  \bibinfo{pages}{182302} (\bibinfo{year}{2014}), \eprint{1404.5963}.

\bibitem[{\citenamefont{Chen et~al.}(2015)\citenamefont{Chen, Son, and
  Stephanov}}]{Chen:2015gta}
\bibinfo{author}{\bibfnamefont{J.-Y.} \bibnamefont{Chen}},
  \bibinfo{author}{\bibfnamefont{D.~T.} \bibnamefont{Son}}, \bibnamefont{and}
  \bibinfo{author}{\bibfnamefont{M.~A.} \bibnamefont{Stephanov}},
  \bibinfo{journal}{Phys. Rev. Lett.} \textbf{\bibinfo{volume}{115}},
  \bibinfo{pages}{021601} (\bibinfo{year}{2015}), \eprint{1502.06966}.

\bibitem[{\citenamefont{Gao et~al.}(2018{\natexlab{b}})\citenamefont{Gao, Pang,
  and Wang}}]{Gao:2018jsi}
\bibinfo{author}{\bibfnamefont{J.-h.} \bibnamefont{Gao}},
  \bibinfo{author}{\bibfnamefont{J.-y.} \bibnamefont{Pang}}, \bibnamefont{and}
  \bibinfo{author}{\bibfnamefont{Q.}~\bibnamefont{Wang}}
  (\bibinfo{year}{2018}{\natexlab{b}}), \eprint{1810.02028}.

\bibitem[{\citenamefont{Liu et~al.}(2019{\natexlab{b}})\citenamefont{Liu, Sun,
  and Ko}}]{Liu:2019krs}
\bibinfo{author}{\bibfnamefont{S.~Y.~F.} \bibnamefont{Liu}},
  \bibinfo{author}{\bibfnamefont{Y.}~\bibnamefont{Sun}}, \bibnamefont{and}
  \bibinfo{author}{\bibfnamefont{C.~M.} \bibnamefont{Ko}}
  (\bibinfo{year}{2019}{\natexlab{b}}), \eprint{1910.06774}.

\bibitem[{\citenamefont{Becattini and Karpenko}(2018)}]{Becattini:2017gcx}
\bibinfo{author}{\bibfnamefont{F.}~\bibnamefont{Becattini}} \bibnamefont{and}
  \bibinfo{author}{\bibfnamefont{I.}~\bibnamefont{Karpenko}},
  \bibinfo{journal}{Phys. Rev. Lett.} \textbf{\bibinfo{volume}{120}},
  \bibinfo{pages}{012302} (\bibinfo{year}{2018}), \eprint{1707.07984}.

\bibitem[{\citenamefont{Adam et~al.}(2019)}]{Adam:2019srw}
\bibinfo{author}{\bibfnamefont{J.}~\bibnamefont{Adam}} \bibnamefont{et~al.}
  (\bibinfo{collaboration}{STAR}), \bibinfo{journal}{Phys. Rev. Lett.}
  \textbf{\bibinfo{volume}{123}}, \bibinfo{pages}{132301}
  (\bibinfo{year}{2019}), \eprint{1905.11917}.

\bibitem[{\citenamefont{Liang and Wang}(2005{\natexlab{b}})}]{Liang:2004xn}
\bibinfo{author}{\bibfnamefont{Z.-T.} \bibnamefont{Liang}} \bibnamefont{and}
  \bibinfo{author}{\bibfnamefont{X.-N.} \bibnamefont{Wang}},
  \bibinfo{journal}{Phys. Lett.} \textbf{\bibinfo{volume}{B629}},
  \bibinfo{pages}{20} (\bibinfo{year}{2005}{\natexlab{b}}),
  \eprint{nucl-th/0411101}.

\bibitem[{\citenamefont{Yang et~al.}(2018)\citenamefont{Yang, Fang, Wang, and
  Wang}}]{Yang:2017sdk}
\bibinfo{author}{\bibfnamefont{Y.-G.} \bibnamefont{Yang}},
  \bibinfo{author}{\bibfnamefont{R.-H.} \bibnamefont{Fang}},
  \bibinfo{author}{\bibfnamefont{Q.}~\bibnamefont{Wang}}, \bibnamefont{and}
  \bibinfo{author}{\bibfnamefont{X.-N.} \bibnamefont{Wang}},
  \bibinfo{journal}{Phys. Rev.} \textbf{\bibinfo{volume}{C97}},
  \bibinfo{pages}{034917} (\bibinfo{year}{2018}), \eprint{1711.06008}.

\bibitem[{\citenamefont{Sheng et~al.}(2019)\citenamefont{Sheng, Oliva, and
  Wang}}]{Sheng:2019kmk}
\bibinfo{author}{\bibfnamefont{X.-L.} \bibnamefont{Sheng}},
  \bibinfo{author}{\bibfnamefont{L.}~\bibnamefont{Oliva}}, \bibnamefont{and}
  \bibinfo{author}{\bibfnamefont{Q.}~\bibnamefont{Wang}}
  (\bibinfo{year}{2019}), \eprint{1910.13684}.

\bibitem[{\citenamefont{{Hess} and {Waldmann}}(1966)}]{Hess:1966}
\bibinfo{author}{\bibfnamefont{S.}~\bibnamefont{{Hess}}} \bibnamefont{and}
  \bibinfo{author}{\bibfnamefont{L.}~\bibnamefont{{Waldmann}}},
  \bibinfo{journal}{Zeitschrift Naturforschung Teil A}
  \textbf{\bibinfo{volume}{21}}, \bibinfo{pages}{1529} (\bibinfo{year}{1966}).

\bibitem[{\citenamefont{{Hess} and {Waldmann}}(1968)}]{Hess:1968}
\bibinfo{author}{\bibfnamefont{S.}~\bibnamefont{{Hess}}} \bibnamefont{and}
  \bibinfo{author}{\bibfnamefont{L.}~\bibnamefont{{Waldmann}}},
  \bibinfo{journal}{Zeitschrift Naturforschung Teil A}
  \textbf{\bibinfo{volume}{23}}, \bibinfo{pages}{1893} (\bibinfo{year}{1968}).

\bibitem[{\citenamefont{Gao and Liang}(2019)}]{Gao:2019znl}
\bibinfo{author}{\bibfnamefont{J.-H.} \bibnamefont{Gao}} \bibnamefont{and}
  \bibinfo{author}{\bibfnamefont{Z.-T.} \bibnamefont{Liang}},
  \bibinfo{journal}{Phys. Rev.} \textbf{\bibinfo{volume}{D100}},
  \bibinfo{pages}{056021} (\bibinfo{year}{2019}), \eprint{1902.06510}.

\bibitem[{\citenamefont{Weickgenannt et~al.}(2019)\citenamefont{Weickgenannt,
  Sheng, Speranza, Wang, and Rischke}}]{Weickgenannt:2019dks}
\bibinfo{author}{\bibfnamefont{N.}~\bibnamefont{Weickgenannt}},
  \bibinfo{author}{\bibfnamefont{X.-L.} \bibnamefont{Sheng}},
  \bibinfo{author}{\bibfnamefont{E.}~\bibnamefont{Speranza}},
  \bibinfo{author}{\bibfnamefont{Q.}~\bibnamefont{Wang}}, \bibnamefont{and}
  \bibinfo{author}{\bibfnamefont{D.~H.} \bibnamefont{Rischke}},
  \bibinfo{journal}{Phys. Rev.} \textbf{\bibinfo{volume}{D100}},
  \bibinfo{pages}{056018} (\bibinfo{year}{2019}), \eprint{1902.06513}.

\bibitem[{\citenamefont{Hattori et~al.}(2019)\citenamefont{Hattori, Hidaka, and
  Yang}}]{Hattori:2019ahi}
\bibinfo{author}{\bibfnamefont{K.}~\bibnamefont{Hattori}},
  \bibinfo{author}{\bibfnamefont{Y.}~\bibnamefont{Hidaka}}, \bibnamefont{and}
  \bibinfo{author}{\bibfnamefont{D.-L.} \bibnamefont{Yang}},
  \bibinfo{journal}{Phys. Rev. D} \textbf{\bibinfo{volume}{100}},
  \bibinfo{pages}{096011} (\bibinfo{year}{2019}), \eprint{1903.01653}.

\bibitem[{\citenamefont{Liu et~al.}(2020)\citenamefont{Liu, Mameda, and
  Huang}}]{Liu:2020flb}
\bibinfo{author}{\bibfnamefont{Y.-C.} \bibnamefont{Liu}},
  \bibinfo{author}{\bibfnamefont{K.}~\bibnamefont{Mameda}}, \bibnamefont{and}
  \bibinfo{author}{\bibfnamefont{X.-G.} \bibnamefont{Huang}}
  (\bibinfo{year}{2020}), \eprint{2002.03753}.

\bibitem[{\citenamefont{Yang et~al.}(2020)\citenamefont{Yang, Hattori, and
  Hidaka}}]{Yang:2020hri}
\bibinfo{author}{\bibfnamefont{D.-L.} \bibnamefont{Yang}},
  \bibinfo{author}{\bibfnamefont{K.}~\bibnamefont{Hattori}}, \bibnamefont{and}
  \bibinfo{author}{\bibfnamefont{Y.}~\bibnamefont{Hidaka}}
  (\bibinfo{year}{2020}), \eprint{2002.02612}.

\bibitem[{\citenamefont{Wang et~al.}(2019)\citenamefont{Wang, Guo, Shi, and
  Zhuang}}]{Wang:2019moi}
\bibinfo{author}{\bibfnamefont{Z.}~\bibnamefont{Wang}},
  \bibinfo{author}{\bibfnamefont{X.}~\bibnamefont{Guo}},
  \bibinfo{author}{\bibfnamefont{S.}~\bibnamefont{Shi}}, \bibnamefont{and}
  \bibinfo{author}{\bibfnamefont{P.}~\bibnamefont{Zhuang}},
  \bibinfo{journal}{Phys. Rev. D} \textbf{\bibinfo{volume}{100}},
  \bibinfo{pages}{014015} (\bibinfo{year}{2019}), \eprint{1903.03461}.

\bibitem[{\citenamefont{Mueller and Venugopalan}(2017)}]{Mueller:2017arw}
\bibinfo{author}{\bibfnamefont{N.}~\bibnamefont{Mueller}} \bibnamefont{and}
  \bibinfo{author}{\bibfnamefont{R.}~\bibnamefont{Venugopalan}},
  \bibinfo{journal}{Phys. Rev.} \textbf{\bibinfo{volume}{D96}},
  \bibinfo{pages}{016023} (\bibinfo{year}{2017}), \eprint{1702.01233}.

\bibitem[{\citenamefont{Mueller and Venugopalan}(2018)}]{Mueller:2017lzw}
\bibinfo{author}{\bibfnamefont{N.}~\bibnamefont{Mueller}} \bibnamefont{and}
  \bibinfo{author}{\bibfnamefont{R.}~\bibnamefont{Venugopalan}},
  \bibinfo{journal}{Phys. Rev.} \textbf{\bibinfo{volume}{D97}},
  \bibinfo{pages}{051901} (\bibinfo{year}{2018}), \eprint{1701.03331}.

\bibitem[{\citenamefont{Wigner}(1939)}]{Wigner:1939cj}
\bibinfo{author}{\bibfnamefont{E.~P.} \bibnamefont{Wigner}},
  \bibinfo{journal}{Annals Math.} \textbf{\bibinfo{volume}{40}},
  \bibinfo{pages}{149} (\bibinfo{year}{1939}), \bibinfo{note}{[Reprint: Nucl.
  Phys. Proc. Suppl.6,9(1989)]}.

\bibitem[{\citenamefont{Kim and Wigner}(1987)}]{Kim:1986gq}
\bibinfo{author}{\bibfnamefont{Y.~S.} \bibnamefont{Kim}} \bibnamefont{and}
  \bibinfo{author}{\bibfnamefont{E.~P.} \bibnamefont{Wigner}},
  \bibinfo{journal}{J. Math. Phys.} \textbf{\bibinfo{volume}{28}},
  \bibinfo{pages}{1175} (\bibinfo{year}{1987}).

\bibitem[{\citenamefont{Kim and Wigner}(1990)}]{Kim:1989wt}
\bibinfo{author}{\bibfnamefont{Y.~S.} \bibnamefont{Kim}} \bibnamefont{and}
  \bibinfo{author}{\bibfnamefont{E.~P.} \bibnamefont{Wigner}},
  \bibinfo{journal}{J. Math. Phys.} \textbf{\bibinfo{volume}{31}},
  \bibinfo{pages}{55} (\bibinfo{year}{1990}).

\bibitem[{\citenamefont{Vasak et~al.}(1987)\citenamefont{Vasak, Gyulassy, and
  Elze}}]{Vasak:1987um}
\bibinfo{author}{\bibfnamefont{D.}~\bibnamefont{Vasak}},
  \bibinfo{author}{\bibfnamefont{M.}~\bibnamefont{Gyulassy}}, \bibnamefont{and}
  \bibinfo{author}{\bibfnamefont{H.~T.} \bibnamefont{Elze}},
  \bibinfo{journal}{Annals Phys.} \textbf{\bibinfo{volume}{173}},
  \bibinfo{pages}{462} (\bibinfo{year}{1987}).

\bibitem[{\citenamefont{Guo}(2020)}]{Guo:2020gxy}
\bibinfo{author}{\bibfnamefont{X.-Y.} \bibnamefont{Guo}},
  \bibinfo{journal}{???} \textbf{\bibinfo{volume}{???}} (\bibinfo{year}{2020}).

\end{thebibliography}

\end{document}